\documentclass[useAMS,usenatbib]{mn2e}
\usepackage{epsfig}
\usepackage{amsmath}
\usepackage{rotating}

\newcommand{\be}{\begin{equation}}
\newcommand{\beq}{\begin{equation}}
\newcommand{\ba}{\begin{eqnarray}}
\newcommand{\ee}{\end{equation}}
\newcommand{\eeq}{\end{equation}}
\newcommand{\ea}{\end{eqnarray}}

\newcommand{\hs}{\hspace{1mm}}

\newcommand{\apj}{ApJ}
\newcommand{\aap}{A\&A}
\newcommand{\apjl}{ApJL}
\newcommand{\mnras}{MNRAS}
\newcommand{\aj}{AJ}

\newcommand{\nat}{{\it Nature}}
\newcommand{\araa}{ARA\&A}

\newcommand{\physrep}{Physics Reports}

\def\lsim{~\rlap{$<$}{\lower 1.0ex\hbox{$\sim$}}}

\def\gsim{~\rlap{$>$}{\lower 1.0ex\hbox{$\sim$}}}

\title[Lyman-Werner Background Fluctuations]{Fluctuations in the High--Redshift Lyman--Werner Background: Close Halo Pairs as the Origin of Supermassive Black Holes}

\author[Dijkstra et al.]{Mark Dijkstra$^{1}$\thanks{E-mail:mdijkstr@cfa.harvard.edu}, Zolt\'{a}n Haiman$^{2}$, Andrei Mesinger$^{3,4}$\thanks{Hubble Fellow}, and J. Stuart B. Wyithe$^{5}$\\
$^{1}$Astronomy Department, Harvard University, 60 Garden Street, Cambridge, MA 02138, USA\\
$^{2}$Astronomy Department, Columbia University, 550 West 120$^{\rm th}$ Street, New York, NY 10027, USA\\
$^{3}$Department of Physics and Astronomy, UCLA, Los Angeles, CA, 90095-1562, USA\\
$^{4}$Department of Astrophysical Sciences, Princeton University, Princeton, NJ 08544, USA\\
$^{5}$School of Physics, University of Melbourne, Parkville, Victoria, 3010, Australia}

\begin{document}

\date{\today}
\pagerange{\pageref{firstpage}--\pageref{lastpage}} \pubyear{2006}

\maketitle

\label{firstpage}
\begin{abstract}
  The earliest generation of stars and black holes must have
   established an early 'Lyman-Werner' background (LWB) at high
  redshift, prior to the epoch of reionization.  Because of the long
  mean free path of photons with energies $h\nu<13.6$eV, the LWB was
  nearly uniform.  However, some variation in the LWB is expected due
  to the discrete nature of the sources, and their highly clustered
  spatial distribution.  In this paper, we compute the probability
  distribution function (PDF)
  of the LW flux that irradiates dark matter (DM) halos collapsing at
  high-redshift ($z\approx 10$). Our model accounts for ({\it i}) the
  clustering of DM halos, ({\it ii}) Poisson fluctuations in the
  number of corresponding star forming galaxies, and ({\it iii})
  scatter in the LW luminosity produced by halos of a given mass
  (calibrated using local observations).  We find that $>99\%$ of the
  DM halos are illuminated by a LW flux within a factor of $2$ of the
  global mean value. However, a small fraction, $\sim
  10^{-8}-10^{-6}$, of DM halos with virial temperatures $T_{\rm
    vir}\gsim 10^4$ K have a close luminous neighbor within $\lsim 10$
  kpc, and are exposed to a LW flux exceeding the global mean by a
  factor of $>20$, or to $J_{\rm 21,LW}>10^3$ (in units of $10^{-21}$
  erg s$^{-1}$ Hz$^{-1}$ sr$^{-1}$ cm$^{-2}$).  This large LW flux can
  photo--dissociate ${\rm H_2}$ molecules in the gas collapsing due to
  atomic cooling in these halos, and prevent its further cooling and
  fragmentation.  Such close halo pairs therefore provide possible sites
  in which primordial gas clouds collapse directly into massive black
  holes ($M_{\rm BH}\approx 10^{4-6}M_{\odot}$), and subsequently grow
  into supermassive ($M_{\rm BH}\gsim 10^9M_{\odot}$) black holes by
  $z\approx 6$.
\end{abstract}

 
\section{Introduction}
\label{sec:intro}

Radiative feedback plays an important role in the formation of stars,
black holes and galaxies. Ionization and heating by photons with
energies exceeding the hydrogen ionization threshold $E> E_{\rm H}$
affect the ability of gas to cool and collapse into dense objects
\citep[e.g][]{CR86,ef92,tw96,K00,feedback}. At the earliest epochs of
structure formation, the dominant coolant of primordial gas clouds at
gas temperatures $T\lsim 10^4$ K are H$_2$ molecules
\citep[e.g.][]{SZ67,LS84,H96}.  The H$_2$ molecules can be
photo--dissociated by ultra--violet (UV) radiation, either directly
(by photons with energies $E>14.7$ eV, if the molecules are exposed to
ionizing radiation) or as a result of electronic excitation by
Lyman--Werner (hereafter LW) photons with energies 11.2eV$\lsim$ E
$<E_{\rm H}$. In this paper we focus on this latter process, which operates
 even in gas that is self-shielded, and/or in gas that is shielded by a neutral intergalactic medium (IGM) prior to the completion of reionization (see \citealt{H97}), from radiation at $E>E_{\rm H}$.

Photodissociation feedback possibly plays an important role in the
formation of the supermassive black holes (SMBHs, $M_{\rm BH}\sim
10^9M_{\odot}$) that existed at redshift $z>6$, when the age of the
universe was $<1$ Gyr \citep[e.g.][]{FanReview}.  Several studies
have modeled the growth of these SMBHs due to accretion and mergers,
starting from stellar--mass seed BHs left behind by the first
generation of stars
\citep[e.g.][]{hl01,h04,ym04,Bromley04,shapiro05,Volonteri06,Li07,JB07,Taka08}. The generic conclusion is that it is challenging to account for the
presence of several$\times10^9~{\rm M_\odot}$ SMBHs by $z\gsim 6$ if
accretion is limited to the Eddington rate, unless the accretion is
essentially uninterrupted over the Hubble time. Gravitational recoil
during mergers can lead to the ejection of growing seed BHs from their
parent halos and can exacerbate the problem
\citep{h04,ym04,shapiro05,Volonteri06,Taka08}. 
An additional difficulty is
that models successfully producing the several$\times10^9~{\rm
  M_\odot}$ SMBHs by $z\gsim 6$ tend to overproduce the number of
lower--mass BHs \citep{Bromley04,Taka08}.

Several alternative mechanisms have been proposed recently that evade
these problems by having a period of rapid super--Eddington growth
\citep[e.g.][]{BL03,Volonteri05,Begelman06,SS06}.  The reason that
photodissociation feedback may provide an interesting formation
channel is that this feedback mechanism can affect gas cooling and
collapse into halos that have $T_{\rm vir}> 10^4$ K
(e.g. \citealt{OH02}). Without H$_2$ molecules, gas inside these halos
collapses nearly isothermally due to atomic cooling, with the
temperature remaining as high as $T\sim 10^4$K, which may strongly
suppress the ability of gas to fragment into stellar mass objects
\citep[e.g.][]{OH02}. Instead of fragmenting, this gas could rapidly
accrete onto a seed BH \citep{Volonteri05}, or collapse directly into
a very massive ($M_{\rm BH}=10^4-10^6M_{\odot}$) black hole
\citep{BL03,K04,Begelman06,Lo06,Lo07,SS06,V08}, possibly with an intermediate state in
the form of a very massive star \citep[see][for a more complete
review]{Omukai08}. If these ``direct-collapse black holes'' indeed
formed, then they would provide a head--start that could help explain
the presence of SMBHs with inferred masses of several $10^9{\rm
  M_\odot}$ per comoving Gpc$^3$ by $z\approx 6$.

Possibly the most stringent requirement for these rapid--growth models
is the absence of H$_2$ molecules during cloud collapse. This requires the existence of a photodissociating background\footnote{Note that the
  gas could remain ${\rm H_2}$--free and close to $T\sim 10^4$K if its
  atomic cooling rate was reduced sufficiently due to the trapping of
  Lyman line radiation \citep{SS06}.}  whose mean intensity exceeds
$J_{\rm 21,LW}\geq J_{\rm crit}\sim 10^3$ \citep{BL03}, where $J_{\rm
  21,LW}$ denotes the intensity in the LW background (in units of
$10^{-21}$ erg s$^{-1}$ Hz$^{-1}$ sr$^{-1}$ cm$^{-2}$, evaluated at
the Lyman limit). Since LW photons propagate nearly unobscured through
the high-redshift (neutral) intergalactic medium until they redshift
into one of the Lyman-series transitions of atomic hydrogen, each halo
simultaneously 'sees' a large number of LW-sources. Each halo is
therefore expected to be exposed to approximately the same LW-flux,
close to the global mean level of the background, which is likely to
be significantly lower than $J_{\rm crit}$ \citep{J08,Omukai08}.

However, some variation in the LW-flux that irradiates individual dark
matter halos is expected. For example, a dark matter halo with a
nearby star forming galaxy (separated, for example, by $d\approx
10$kpc) will be exposed to a LW flux exceeding the global value, and
reaching $J_{\rm 21,LW}\sim 10^3(d/10\hs{\rm
  kpc})^{-2}[\dot{M_*}/(20M_{\odot}\hs {\rm yr}^{-1})]$, where
$\dot{M_*}$ is the star formation rate in the nearby galaxy (see
\S~\ref{sec:uvpdf} below for a more detailed discussion). In the
present paper, we model this variation and compute the probability
distribution function (PDF) of the LW-flux that irradiates 
high-redshift dark matter halos.  In particular, we focus on the
high--flux tail of this PDF, as our goal is to answer the following
questions: what fraction of DM halos with $T_{\rm vir}> 10^4$ K,
collapsing at high redshift, are exposed to a flux exceeding $J_{\rm
  crit}$?  Is this fraction sufficient to account for the space
density of several$\times10^9~{\rm M_\odot}$ SMBHs at $z\approx 6$ 
inferred from observations?

In the model we present below, we account for the clustering of dark
matter halos. Proper modeling of clustering is important when one
wants to model the high LW-flux tail of the distribution, which is
dominated by close pairs of halos. We also account for Poisson
fluctuations in the number of star forming galaxies surrounding a
halo, and allow for a scatter in LW luminosity produced in a halo with
a given mass. In the local universe, the UV luminosity can vary by
orders of magnitude for a given stellar mass \citep[e.g][]{DS07}, and
a similar variation of the UV luminosity is expected for a given total
halo mass.

Spatial variations of the UV background have been investigated
previously by several authors, especially in the context of the Lyman
$\alpha$ forest at lower--redshift. For example, \citet{Zuo92} studied
the expected fluctuations in the {\it ionizing} radiation field produced by
randomly distributed sources, and showed that the variance is
significantly increased due to radiative transfer effects. More recently,
\citet{MD08} studied the expected fluctuations of the ionizing background 
during the advanced stages of reionization using 'semi-numerical' simulations.
Previous works have studied various observable consequences of
fluctuations in the ionizing background on the lower redshift ($z\lsim
5.5$) Ly$\alpha$ forest \citep[e.g.][]{Croft04,Meiksin04,Schaye06}. Existing work on the evolution of the LW background has focused on its global (volume averaged) build-up with time, and the impact on subsequent star formation \citep[e.g.][]{Haiman00,M06,J08}.The main differences of our study from these earlier works are that (i) we include the non--linear clustering of sources, (ii) we
specialize to compute the PDF of the flux as seen by high--redshift DM
halos, (iii) we allow the UV luminosity for a given halo mass to be variable, instead of assuming a rigid one-to-one correspondence between halo mass and UV luminosity, (iv) we focus on the tail of the flux PDF, and (v) we discuss the significance of this tail for rapid high--redshift SMBH growth (note however that the work presented by Mesinger \& Dijkstra (2008) did include effects (i) and (ii)). While this paper was being completed, we became aware of related work
by \citet{Ahn08}, who studied the inhomogeneity of the LW
background with cosmological simulations. Their paper focuses mostly
on the impact of these fluctuations on the star formation efficiency
in minihalos (as opposed to the tail of the PDF and its significance
to SMBH formation, which is the focus of our study). The papers seem
to agree wherever there is overlap: the $z=10.5$ LW-flux PDF shown in
Fig.~11 of \citet{Ahn08} is consistent with our results (e.g. our
Fig.~\ref{fig:pdf}).

The outline of this paper is as follows:
In \S~\ref{sec:model}, we describe our model for calculating the LW flux PDF. 
In \S~\ref{sec:results}, we present our results. 
Model uncertainties and the implications of our work are discussed in \S~\ref{sec:discuss}, before summarizing our main conclusions in \S~\ref{sec:conc}.
The parameters for the background cosmology used throughout this paper are $(\Omega_m,\Omega_{\Lambda},\Omega_b,h,\sigma_8)=(0.27,0.73,0.042,0.70,0.82)$, consistent with 5--year data from the {\it Wilkinson Microwave Anisotropy Probe (WMAP)} \citep{wmap5,wmap}.

\section{The Model}
\label{sec:model}

This section, which describes our numerical modeling, is split into
three parts. In \S~\ref{sec:sec1}, we describe how we model the
non--linear spatial clustering of sources around the position of a
dark matter halo. In \S~\ref{sec:mc}, we describe how we generate a
discrete number of UV sources from a continuous density field, and how
we assign UV-luminosities to individual sources by using a Monte-Carlo
(MC) technique. In \S~\ref{sec:sec3}, we discuss how we finally obtain
the total intensity of the Lyman-Werner radiation field ($J_{\rm LW}$)
that the halo is exposed to.  This procedure yields individual
realizations for $J_{\rm LW}$ seen by a single halo. A distribution
for $J_{\rm LW}$ is obtained by performing multiple Monte-Carlo realizations
for a range of halo masses.

\subsection{Modeling the Lyman-Werner Background as Sampled by Dark Matter Halos}
\label{sec:sec1}

\subsubsection{The Clustering of Sources around a Halo}
\label{sec:clustering}

Consider a halo of total (dark matter+gas) mass $M$ that virializes at
redshift $z$. The average number $\mathcal{N}(m,r)dmdr$ of halos within
the mass range $m\pm
dm/2$ that populate a surrounding
spherical shell of physical radius $r$ and thickness $dr$, is given by
\begin{eqnarray}
\nonumber
\mathcal{N}(m,r)dmdr&=&4\pi r^2 dr \\ 
&&\hspace{-15mm}\times (1+z)^3 \frac{dn_{\rm ST}(m,z)}{dm}dm\Big{[}1+\xi(M,m,z,r)\Big{]}.
\label{eq:fmr}
\end{eqnarray}
Here ${dn_{\rm ST}(m,z)}/{dm}$ is the Press-Schechter (1974) mass
function \citep[with the modification of][]{ST01}, which gives the
number density of halos of mass $m$ (in units of {\it comoving}
Mpc$^{-3}$). The factor $(1+z)^3$ converts the number density of halos
into proper Mpc$^{-3}$.

For simplicity, we assume a static Euclidean space in our
calculations.  Cosmological corrections affect only sources at
distances comparable to the Hubble length from the central halo, and
we have explicitly verified that the value of the global LW-background
would be changed by less than a factor of $\sim 2$. Since this is
within the uncertainty in the flux attenuation due to the uncertain
intergalactic H$_2$ abundance (see \S~\ref{sec:lwrt}), we have ignored
these cosmological corrections throughout our paper for simplicity.

The quantity $\xi(M,m,z,r)$ denotes the two-point correlation
function, which gives the excess (above random) probability of finding
a halo of mass $m$ at a distance $r$ from the central halo. In this
paper we are especially interested in the high--end tail of the PDF
flux impinging upon halos of mass $M$. Since, as we demonstrate below,
this tail is dominated by close pairs of halos, we would like to model
the clustering of closely separated halos as accurately as
possible. To this end we use the analytic formulation of non-linear
Eulerian bias developed by \citet{Iliev03}, which fits the two-point
correlation function derived from N-body simulations significantly
better than the standard linear--bias approximation
\citep[][]{MoWhite96,SB02}, especially at small
separations.

\begin{figure}
\vbox{\centerline{\epsfig{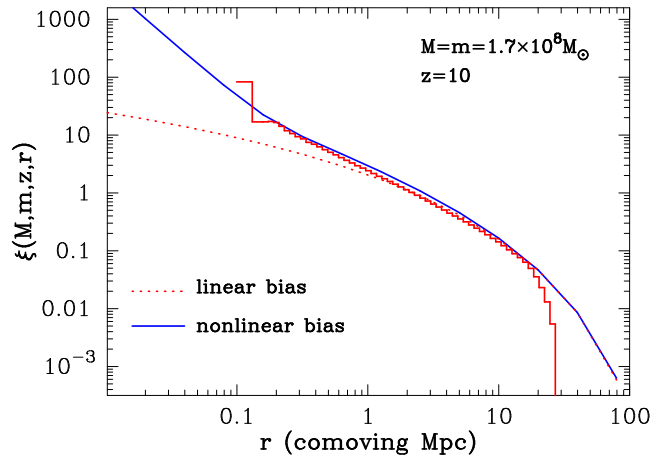}}}
\caption[]{The two-point correlation function $\xi(M,m,z,r)$ is shown
  for redshift $z=10$, and halo masses $M=m=1.7 \times 10^8 M_{\odot}$. The solid (dashed)
  curve shows $\xi(M,m,z,r)$ obtained using the non-linear (linear)
  bias approximation, while the histogram shows the two--point
  correlation function derived from a semi--numerical simulation (see
  text). The non--linear bias formalism matches the simulation better
  (as shown by \citealt{Iliev03}), and therefore provides a more
  accurate description of clustering of closely separated halos. Note
  that in the range of $r\approx 10-100$ (comoving) kpc, the two-point
  function in the non-linear case exceeds the linear--bias prediction
  by 1-2 orders of magnitude.}
\label{fig:2pt}
\end{figure}

The benefit of using this non-linear bias formalism is illustrated in
Figure~\ref{fig:2pt}, where we show, as an example, the two-point
correlation function $\xi(M,m,z,r)$ for $z=10$, and $M=m=1.7 \times
10^8 M_{\odot}$.  Here, the solid (dashed) curve shows the non-linear
(linear) bias approximation, while the histogram shows the two--point
correlation function derived from a 'semi-numerical' simulation
\citep{MF07}. In this simulation, an excursion-set approach is
combined with first-order Lagrangian perturbation theory to generate
density, velocity, and halo fields at $z=10$ (for a more detailed
description of this approach, the reader is referred to 
\citealt{MF07}).
The simulation does not resolve scales below $r \sim0.14$ Mpc. Clearly, the non-linear bias formalism matches the
simulation much better, especially at $r< 1$ (comoving)
Mpc.
Furthermore, Figure~\ref{fig:2pt} shows
that in the range of $r\approx 10-100$ (comoving) kpc, the two-point
function in the non-linear case exceeds the linear--bias prediction by
1-2 orders of magnitude. Recently, \citet{Hennawi} have found a large excess of close quasar pairs, which implies a steepening of the (projected) two-point correlation function, which is perhaps an observational signature of nonlinear clustering.

The ionizing and Lyman-Werner flux received from a halo of mass $m$ at
a distance $r$ depends on the UV-emissivity of this particular halo,
which is likely to be variable (while this paper was reviewed, Lee et al. 2008 posted a paper that presents a detailed study of the variation of UV luminosity with halo mass, as we discuss next.

\subsubsection{The UV Luminosity of Dark Matter Halos}
\label{sec:uvpdf}

For the UV luminosity of a dark matter halo, we adopt the empirical
relation $L_{\rm LW}\sim 8 \times 10^{27}\dot{M}_*$ erg s$^{-1}$
Hz$^{-1}$ \citep{K98}, where $\dot{M}_*$ is the star formation rate
inside the halo in units of $M_{\odot}$/yr.

The star--formation rate in dark matter halos is highly uncertain
(one of the reasons being radiative feedback itself), especially at
the redshifts of interest ($z\sim 10$) where essentially no data
exists. The best constraints on the efficiency of star formation in
the high--redshift Universe are provided by observations of $z\sim 6$
galaxies, which populate a Universe that is ``only'' 0.5 Gyr older
than at $z\sim 10$. Among observed galaxy samples, the $z\sim6$
population is therefore likely to bear closest resemblance to the
$z\sim 10$ population. Note that \citet{Stark07} have found a few
gravitationally lensed emission--line objects, which provide possible
candidates for $z=10$ Ly$\alpha$ emitting galaxies. The small number
of candidates, and the small volume that they populate, prohibit
strong constraints on the global star formation efficiency at
$z=10$. However, if the candidates are indeed  $z=10$ Ly$\alpha$ emitters,
then their abundance would imply that star formation is very efficient at $z=10$ (see Fig~6 of Mesinger \& Furlanetto, 2008), which would render
our estimates below of the fraction of halos that ``sees'' an
anomalously large $J_{\rm LW}$ conservative.

One can reproduce the observed rest--frame UV luminosity function of
Lyman Break Galaxies at $z\sim 6$ in the Hubble Deep Fields
\citep{Bouwens06} by simply assuming that a fraction $\epsilon_{\rm
  DC}\sim 0.1$ of all dark matter halos are converting $f_*\sim 0.1$
of their gas into stars over their duty--cycle of $\epsilon_{\rm
  DC}t_{\rm hub}\sim 0.1t_{\rm hub}$, where $t_{\rm hub}$ is the
Hubble time \citep{WL06}. Furthermore, such a model also successfully
reproduces the observed luminosity functions of Ly$\alpha$ emitting
galaxies at $z=5.7$ and $z=6.5$ \citep[e.g][]{LF}, which suggests it
provides a decent prescription of star formation at $z\sim 6$. If we
apply the same star formation efficiency to the halos of interest,
then we find that
\begin{equation}
\langle L_{\rm LW,26}\rangle(m)=2.8\Big{(}\frac{m}{10^8 M_{\odot}}\Big{)}\Big{(}\frac{1+z}{11.0}\Big{)}^{-1.5}, 
\label{eq:epsion}
\end{equation} 
where the UV-luminosity is in units of $10^{26} \hs {\rm erg} \hs {\rm s}^{-1} \hs {\rm Hz}^{-1}$. 

Equation~(\ref{eq:epsion}) implies a one-to-one relation between the
mass $m$ of a dark matter halo and its UV-emissivity. However, in the
local universe the UV luminosity varies by orders of magnitude for a
given stellar mass \citep[e.g. Fig.~7 of][]{DS07}.

In this paper, we model the spread in UV luminosity produced in halos
with halos of mass $m$ with a log--normal distribution,
$P(L_{\rm LW,26},m)$. The quantity $L_{\rm LW,26}$ here
denotes the LW-luminosity per unit star formation rate, in units of
$10^{26}$ erg s$^{-1}$ Hz$^{-1}$.  The
probability that a halo of mass $m$ has a UV luminosity in the range
$\log L_x \pm d\log L_x/2$ ('x' denotes 'LW,26') is given by

\begin{eqnarray}
\nonumber
P(\log L_x,m)d\log L_x&=&\\  
&&\hspace{-25mm}\frac{d \log L_x}{\sigma_x\sqrt{2\pi}}\exp \Big{[}\frac{-(\log L_x- \log [\mathcal{F} \times \langle L_x \rangle])^2}{2\sigma_x ^2}\Big{]},
\label{eq:uvpdf}
\end{eqnarray} 
where log denotes log$_{10}$. In this paper, our fiducial model adopts
$\sigma_{\rm LW}=0.50$. Hence, the Lyman-Werner flux is larger/smaller
by an order of magnitude than the value given by
equation~(\ref{eq:epsion}) for 2.5\% of the halos (for all $m$). The
factor $\mathcal{F}$ reduces the mean luminosity $\langle L_x
\rangle$ relative to the no--scatter prediction in
equation~\ref{eq:epsion}.  Since the inclusion of scatter increases
the predicted abundance at fixed UV luminosity, this factor is
necessary to maintain consistency between our model and the observed
$z=6$ rest-frame UV luminosity function \citep{Bouwens06}.  In our
fiducial model, we find $\mathcal{F}=0.38$; increasing $\sigma_{\rm
  LW}$ requires lower values of $\mathcal{F}$ (see
Table~\ref{table:sigma}).  We emphasize that the distribution given by
equation~\ref{eq:uvpdf} is clearly ad--hoc -- it is meant to crudely
represent the luminosity scatter one may reasonably expect among high-redshift
halos. The impact of varying the parameter $\sigma_{\rm LW}$ is
discussed in \S~\ref{sec:param}.
\begin{table}
\centering
\caption{The relation between the parameters $\sigma_{\rm LW}$ and $\mathcal{F}$ that are used to model the scatter 
  in UV luminosity for a given halo mass (see eq.~\ref{eq:uvpdf}).}
\begin{tabular}{c c c c c }
\hline\hline
$\sigma_{\rm LW}$ & $0.0$ & $0.25$ & $0.50$ & $0.66$ \\
$\mathcal{F}$ & $1.0$ &$0.88$& $0.38$& $0.21$\\
\hline
\end{tabular}
\label{table:sigma}
\end{table}

\subsection{Monte-Carlo Realizations of the Model}
\label{sec:mc}

The environment of a halo is sampled by $N_{r}$ concentric spherical
shells from $r=r_{\rm min}$ out to a maximum radius $r_{\rm max}$.  We
denote the radius and thickness of shell number $j$ by $r_j$ and
$dr_j$, respectively. Furthermore, the mass function ${dn_{\rm
    ST}(m,z)}/{dm}$ is sampled by $N_m$ mass bins that are spaced
evenly in $\log m$. Mass bin number $i$ contains halos in the mass
range $\log m_i\pm d\log m_i/2$.

\subsubsection{The Clustering of Sources around a Halo}
\label{sec:nmi}
Equation~(\ref{eq:fmr}) gives the average number of halos
$\mathcal{N}(m,r)dmdr$ within the mass range $m\pm dm/2$ and within the
spherical shell at $r \pm dr/2$. The actual number $N(m_i,r_j)$ in
shell number $j$ in the $i$th mass bin  fluctuates around
this average, and in the Monte-Carlo simulation we populate this shell
with halos by drawing from a Poisson distribution with a mean
$\langle N \rangle_{\rm i,j}\equiv \mathcal{N}(m_i,r_j)dm_idr_j$. 

\subsubsection{The UV Luminosity of the Surrounding Halos}

As mentioned above, only a fraction $\epsilon_{\rm DC}\sim 0.1$ of all
halos are actively forming stars at a given time. We determine whether
a halo is ``on'' or ``off'' by generating a random number $0\leq R\leq
1$.  If $R \leq \epsilon_{\rm DC}$, then the halo is ``on'' (i.e. star
forming) and vice versa. For each star--forming halo that occupies
radial shell number $j$ and mass bin number $i$, we generate UV
luminosities $L_x(m_i)$ from the probability distribution given
in equation~(\ref{eq:uvpdf}) above.  Specifically, we convert
uniformly distributed random numbers $0\leq R\leq 1$ to luminosities
$\log L_x(m_i)$ using
\begin{equation}
R=\int_{-\infty}^{\log L_x(m_i)}P(u,m_i)du.
\end{equation} 
This equation can be inverted analytically and yields $\log
L_x(m_i)=\log [\mathcal{F}
\times\langle L_x\rangle(m_i)]+\sigma_x\sqrt{2}{\rm
  erf}^{-1}(2R-1)$, where erf$^{-1}$ is the inverse error function,
and $\langle L_x\rangle(m)$ is given by
equation~(\ref{eq:epsion}).

\subsection{The UV-flux at the Halo}
\label{sec:sec3}

Using the prescription of \S~\ref{sec:mc} for populating the
environment of a (single) halo with UV-emitting galaxies, we can
calculate the Lyman-Werner flux that is ``seen'' by the halo from
\begin{equation}
J_{\rm LW}=\frac{1}{4\pi}\sum_{i=1}^{N_{\rm m}}\sum_{j=1}^{N_{\rm r}} \sum_{k=1}^{N(m_i,r_j)}\frac {L_{\rm LW,k}(m_i)}{4\pi r^2_j}\times \Delta(R-\epsilon_{\rm DC,k}).\label{eq:lw}
\end{equation}
Here $\Delta(x)$ is the Heavyside step function, and the factor
$1/4\pi$ preceding the summation symbols converts the units of $J_{\rm
  LW}$ into erg s$^{-1}$ cm$^{-2}$ Hz$^{-1}$ sr$^{-1}$ (the units of
$L_{\rm LW}$ are erg s$^{-1}$ Hz$^{-1}$). The sum over $k$
reflects the fact that any given mass-radius bin may contain $>1$
halos. For sufficiently large radii and low masses, we often have
$\langle N \rangle_{\rm i,j}\gg 1$. In this case, we do not generate UV luminosities
for individual halos. Instead, whenever $\langle N \rangle_{\rm i,j} \geq 10$, the
total UV luminosity from that shell is given by its mean value,
i.e. $L_{\rm x,i,j}=\langle N \rangle_{\rm i,j}\times \epsilon_{\rm DC}\times
\langle L_{\rm x} \rangle (m_i)$. This approximation speeds up
the calculation, and does not affect the $J_{\rm LW}$ PDF in the tail.
The main reason is simple: the high-$J_{\rm LW}$ tail of the PDF
arises from Poisson fluctuations in the number of halos in the inner
radial shells, and from fluctuations in the UV-luminosity of these
sources (see below).

Equation~(\ref{eq:lw}) gives the Lyman-Werner flux that is
seen by a single halo. We repeat the Monte-Carlo calculation $N_{\rm
  mc}$ times in order to derive an accurate PDF of $J_{\rm LW}$. We would
like to investigate whether halos exist that are irradiated by a
LW-background that exceeds $J_{\rm crit}=10^3$, so that the gas inside
these halos may collapse directly into a massive black hole. This
scenario was introduced as a possible formation mechanism for the
seeds of $z=6$ supermassive black holes (see
\S~\ref{sec:intro}). These black holes are extremely rare, with a
number density of $\sim 1$ (comoving) Gpc$^{-3}$ (e.g. Fan et
al. 2001, 2003). Hence (approximately) only one halo per cGpc$^3$
needs to yield a seed. For comparison, there are $\sim 10^9$ halos at
$z=10$ with virial temperatures exceeding $T_{\rm vir}=10^4$ K in a
$1$ cGpc$^3$ volume. Ideally, we would therefore want to perform
$N_{\rm MC}\sim 10^9$ Monte-Carlo simulations, and verify whether at
least one halo is exposed to a LW-flux that exceeds $J_{\rm
  crit}$. However, this is computationally prohibitively expensive. In
practice, we instead utilize $N_{\rm MC}\sim 10^7$ realizations, and
complement our Monte-Carlo simulations with analytic calculations as
described in \S~\ref{sec:param}.

\section{Results}
\label{sec:results}

We use $N_m=500$ and $N_r=100$ throughout this paper. We have verified
that our results are not sensitive to the precise choices of $N_m$ and
$N_r$.

\begin{table}
\centering
\caption{Fiducial model parameters (see \S~\ref{subsec:modelparams} for discussion).}
\begin{tabular}{c c c c c c}
\hline\hline
M & $m_{\rm min}$ & $r_{\rm min}$  & $r_{\rm max}$ & $\sigma_{\rm LW}$ & $N_{\rm mc}$\\
($M_{\odot}$) &($M_{\odot}$) &  (pkpc)& (pMpc) & &\\
\hline
 $4\times 10^7$ &$4\times 10^7$ & $2r_{\rm vir}=2.0$ & 18 & 0.50 & $10^7$\\
\hline\hline
\end{tabular}
\label{table:model}
\end{table}

\subsection{The Model Parameters}
\label{subsec:modelparams}

In our fiducial model, we investigate the flux PDF seen by halos of
mass $M=M_H=4\times 10^7$ M$_{\odot}$ at $z=10$. These halos are just
massive enough to excite atomic H cooling processes (i.e. T$_{\rm vir}=10^4$K, e.g. eq.~25 in Barkana \& Loeb 2001, for a mean molecular weight $\mu=1.2$).  As mentioned above, the number density of sources more massive than $M_H$ at $z=10$
is $\sim 1$ cMpc$^{-3}$. The impact of varying $M$ is investigated in \S~\ref{sec:param}.

\begin{figure}
\vbox{\centerline{\epsfig{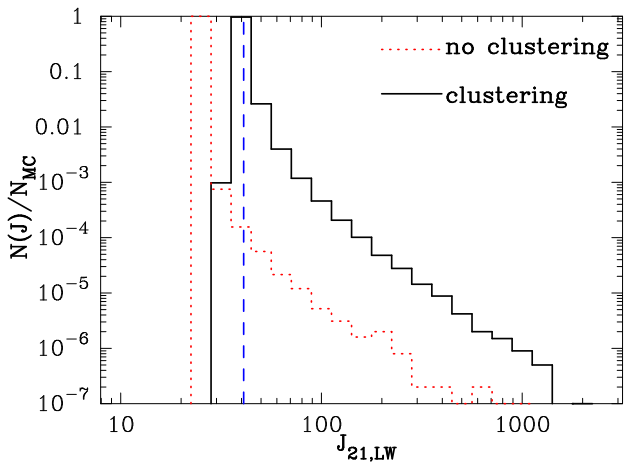}}}
\caption[]{The probability distribution (PDF) of the Lyman-Werner flux
  $J_{\rm LW}$ (in units of $J_{\rm 21,LW}=10^{-21}$ erg s$^{-1}$
  cm$^{-2}$ sr$^{-1}$ Hz$^{-1}$) as seen by halos of mass $M=4\times
  10^7$ M$_{\odot}$ at $z=10$. The {\it black solid (red dotted)} histogram
  corresponds to our fiducial model with (without) clustering. The
  dashed vertical line at $J_{\rm 21,LW}=40$ denotes the mean value of
  the LW background (see text). The figure shows that ({\it i}) the
  vast majority of halos see a LW flux that is within a factor of $2$
  of the mean value, ({\it ii}) the clustering of halos boosts the
  tail of the flux PDF by more than an order of magnitude, and ({\it iii}) an
  exceedingly small, but non--zero fraction ($\sim$ few $\times
  10^{-7}$) of dark matter halos see a LW flux boosted to levels
  exceeding $J_{\rm 21,LW}>J_{\rm crit}=10^3$.}
\label{fig:pdf}
\end{figure}

\begin{figure*}
\vbox{\centerline{\epsfig{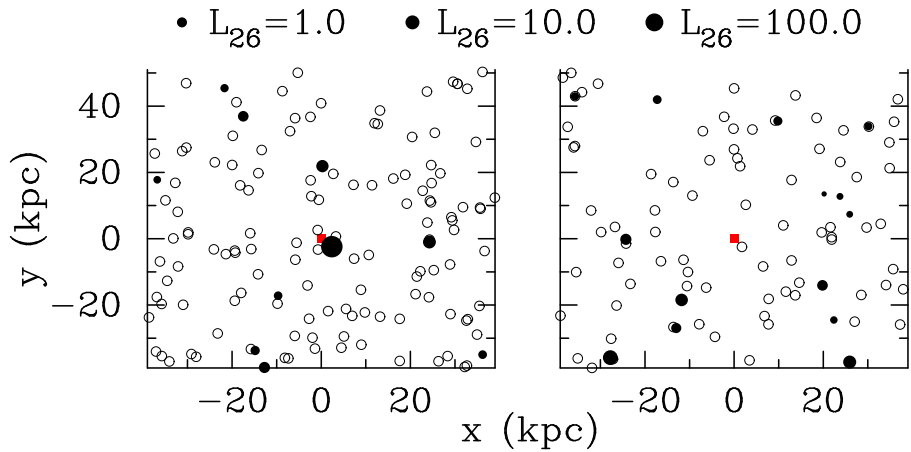}}}
\caption[]{Two realizations of the environment of a fiducial
  $M=4\times 10^7$ M$_{\odot}$ halo at $z=10$.  The {\it left} ({\it
    right}) {\it panel} shows a halo that ``sees'' a $J_{21,{\rm LW}}=10^{3}$ ($J_{21,{\rm LW}}=40$).  The fiducial halo is
  represented by the {\it filled square} at $x=y=0$ kpc. {\it Open
    circles} denote surrounding galaxies that are not actively forming
  stars, and therefore not contributing to the local LW flux. {\it
    Solid circles} denote halos that are actively forming stars; their
  UV luminosity is represented by the size of the circle (the largest
  circles are the most luminous, as illustrated by the graphic scale
  above the panels). The halo that sees $J_{21,{\rm LW}}\sim 10^{3}$
  has a nearby ($r\sim 3$ kpc) neighbor that is luminous
  ($L_{\rm LW,26}=10^2$), while the halo in the {\it right
    panel} has no luminous nearby neighbor, and sees $J_{21,{\rm LW}}\sim 40$. The latter configuration is most common.}
\label{fig:diagram}
\end{figure*} 

The minimum distance of the center of the halo to the center of
another nearby halo, $r_{\rm min}$, has to be specified in our
model. In the absence of peculiar velocities, one may expect $r_{\rm
  min}>r_{\rm vir,M}$. However, close pairs of halos can merge, and
share a common halo, making $r_{\rm min}<r_{\rm vir,M}$ possible. On
the other hand, (very) close halo pairs are short-lived, and will
merge on a time scale shorter than $\epsilon_{\rm DC}t_{\rm hub}$. The
precise choice of $r_{\rm min}$ is therefore somewhat arbitrary; in
the fiducial model we adopt $r_{\rm min}=2r_{\rm vir}=2.0$ kpc, and we will
investigate the impact of varying $r_{\rm min}$ in \S~\ref{sec:param}.

In our calculation, we only consider halos out to $r_{\rm max}=(\lambda_{\rm LW,1}-\lambda_{\beta})c/[\lambda_{\beta}H(z)]\sim 18$ pMpc, in which $\lambda_{\rm LW,1}=1110$ \AA\hs is the wavelength corresponding to the lowest energy Lyman-Werner transition. Lyman-Werner photons that were emitted at $r>r_{\rm max}$ have redshifted into one of the Lyman-series resonances prior to
reaching the halo, and are converted efficiently into photons with
energies $<10.2$ eV before reaching our halo. In reality, 18 pMpc is
an upper limit, and $r_{\rm max}$ should be a strong function of
frequency \citep{H97}. The impact of LW radiative transfer on our
results is discussed further in \S~\ref{sec:lwrt}; in practice, the
high--flux tail of the PDF is insensitive to the choice of $r_{\rm max}$.

Finally, we assume that the minimum mass for star forming halos is
$m_{\rm min}=4\times 10^7$ M$_{\odot}$. This model therefore
represents a universe in which no star--formation occurs in
``minihalos'' with T$_{\rm vir}=10^4$ K (e.g. due to the presence of
the LW background itself or due to an X-ray background). The impact of
varying $m_{\rm min}$ is investigated in \S~\ref{sec:param}. The
parameters of our fiducial model are summarized in
Table~\ref{table:model}.

\subsection{The Flux PDF in the Fiducial Model}

In Figure~\ref{fig:pdf}, we show our main results: histograms showing
the fraction of Monte-Carlo realizations that yield a Lyman-Werner
flux $J_{\rm LW}$ as a function of $J_{\rm LW}$. The solid histogram
shows the fiducial model. The distribution in $J_{\rm LW}$ peaks
at around $J_{\rm 21,LW}=40$, and $95\%$ of the models lie in the range
$35\lsim J_{\rm 21,LW}\lsim 45$. The distribution is highly asymmetric
with a long tail towards high values of $J_{\rm 21,LW}$.

The sharp cut-off on the low-$J_{\rm 21,LW}$ side is due to the fact that
the IGM is assumed to be transparent to LW-photons, which causes a large number  of distant sources to contribute to the intensity of the local LW
radiation field (for reference, the total number of halos with $m>m_{\rm min}$ within a sphere of radius $r_{\rm max}=18$ pMpc is $\gsim 10^7$!). Poisson fluctuations in a large number of distant
sources, as well as variations in their UV luminosity, only result in
a small variation in $J_{\rm LW}$. Indeed, the dashed vertical line at
$J_{\rm 21,LW}=40$ corresponds to the value of $J_{\rm LW}$ in the
absence of any scatter in UV luminosity of individual halos
(i.e. $\sigma_{\rm LW}=0.0$), and with no Poisson fluctuations in the
number of halos. This can be thought of as the level of the global
mean LW background. No halos see a flux that is significantly smaller
than this mean background value of $J_{\rm 21,LW}=40$ (i.e. among our
$10^7$ realizations, and with our flux--bin resolution, there are no
instances when the flux falls below the lowest flux--bin. Note that this value differs for the runs that do and do not include clustering).

On the other hand, the long tail towards high-$J_{\rm 21,LW}$ arises
from Poisson fluctuations in the number of close halos, and from the
assumed scatter in their UV luminosity. The highest values of $J_{\rm
  21,LW}$ are seen by halos that have a luminous nearby neighbor. This
is illustrated graphically in the {\it left panel} of
Figure~\ref{fig:diagram}, where we show the environment of a halo that
is irradiated by a LW-flux close to the maximum value we find, $J_{\rm
  21,LW}\sim10^3$. The small {\it (red) square} at $x=y=0$ kpc denotes
the position of the central halo, while {\it open circles} denote
halos in which no star formation occurs, and are therefore not producing
any LW flux. The {\it filled circles} denote halos that are forming
stars, with the magnitude of the UV-luminosity represented by the size
of the circles. For reference, the sizes corresponding to
$L_{\rm LW,26}=1.0,10.0$ and $100.0$ are shown as labeled above
the figure panels. The central halo is illuminated by a bright
($L_{\rm LW,26}\sim 10^2$), nearby ($r\approx 3$ kpc) star
forming halo.\footnote{Our spherically symmetric clustering model
  generates only radial distances; we have artificially generated
  random azimuthal angles to place the halos in the $x-y$ plane in
  Figure~\ref{fig:diagram}. Hence, Figure~\ref{fig:diagram} does not
  represent an accurate projection or slice of a realistic 3D source
  distribution; it is meant only to graphically illustrate the
  difference in the environment of halos illuminated by a high versus
  a low $J_{\rm LW}$.}  For comparison, the {\it right panel} shows
the environment of a much more common halo that sees $J_{\rm
  LW,21}=40$. Here, the nearest source is located at a distance of
$r=25$ kpc and has a UV luminosity of $L_{\rm LW,26}=10$ (this source only contributes $\sim 2\%$ of the total LW flux
seen by the central halo at $r=0$). 

Note that this figure also illustrates the importance of
clustering. Figure~\ref{fig:diagram} shows the distribution of halos
in a surrounding volume that is $\sim 0.35$ comoving Mpc$^3$; without
clustering, according to the Sheth-Tormen mass function, this volume
would contain, on average, only $\sim 1.4$ dark matter halos with $T_{\rm vir}\gsim
10^4$K.  In the presence of clustering, we find several tens of halos
in both realizations shown in the two panels.

Figure~\ref{fig:pdf} shows that even though the distribution of
$J_{\rm LW}$ drops off rapidly with increasing $J_{\rm LW}$, the
fraction few $\times 10^{-7}$ of all halos see $J_{\rm
  21,LW}\geq 10^3$. This corresponds to $10^3$ halos in a volume of
1 Gpc$^{3}$. Note that because the high--$J_{\rm LW}$ tail is dominated
by close pairs of halos, accounting for clustering is quite important
in the present calculation. This is quantified by the {\it red dotted
  curve} which shows the fiducial model, but without source clustering
(i.e. $\xi(M,m,z,r)=0$ in eq.~\ref{eq:fmr}). In the absence of
clustering the fraction of halos that sees a given $J_{\rm LW}$ above
the mean value is typically smaller by about 1.5 orders of magnitude\footnote{The enhancement of the high-flux tail of the PDF due to clustering varies only weakly with $J_{\rm 21,LW}$. The reason for this weak dependence is that especially the tail-end of the PDF at $J_{\rm LW,21}\gsim 200$ arises mainly from single UV-sources that are in close proximity ($2$ kpc$<r<7$ kpc, see \S~\ref{sec:param}) to the central halo. Over this narrow range of separations, the two-point function $\xi(M,m,z,r)$ -- and hence the expected number of surrounding halos -- varies relatively weakly (the clustering length for halos of mass $M$ is $\sim 300$ pkpc, see Fig.~\ref{fig:2pt})}.

Our main result, presented in this section, is that a non-negligible
fraction of halos, i.e. a few $\times 10^{-7}$, may see a local LW
flux that is higher by almost two orders of magnitude compared to the
mean global background (corresponding to a model in which Poisson
fluctuations in the number of sources, their clustering, and
variations in their UV-luminosity are ignored). In the next section,
we investigate the sensitivity of this result to the assumed model
parameters.

\subsection{The Impact of Model-Parameter Variations}
\label{sec:param}

The impact of varying the model parameters on the very high $J_{\rm
  LW}$ tail of the flux distribution can be efficiently investigated
using an analytic approach. This is because the high--$J_{\rm LW}$
tail is dominated by halos with a nearby star--forming halo, and, in
this limit, the
Poisson probability of having two or more such nearby halos is
negligibly small. The probability that a halo has {\it at least one}
neighbor that would alone expose it to a
Lyman-Werner flux in the range $\log J_{\rm LW}\pm d \log J_{\rm
  LW}/2$, is given by

\begin{eqnarray}
\nonumber
P_1(J_{\rm LW})d\log J_{\rm LW}&=&d\log J_{\rm LW}\times \epsilon_{\rm DC} \times \\ 
&&\hspace{-27mm}\sum_{i=i_{\rm min}}^{N_{\rm m}}\sum_{j=j_{\rm min}}^{N_{\rm r}}\langle N \rangle_{\rm i,j}\hs{\rm e}^{-\langle N \rangle_{\rm i,j}}P(\log L[J_{\rm LW},r_j],L_{x}[m_i]) 
\label{eq:analytic}
\end{eqnarray} 
where $\langle N \rangle_{\rm i,j}=\mathcal{N}(m_i,r_j)dm_idr_j$ (see \S~\ref{sec:nmi}), $r(i_{\rm min})=r_{\rm min}$, $m(j_{\rm min})=m_{\rm min}$, and
$P(\log L[J_{\rm LW},r],L_{x}[m])\hs d\log$ $J_{\rm LW}$
denotes the probability that a halo of mass $m$ that is located at a
distance $r$ produces a LW luminosity in the range $\log J_{\rm LW}\pm
d \log J_{\rm LW}/2$. This halo must have a Lyman-Werner luminosity
that is $L=16\pi^2 r^2J_{\rm LW}$ (in erg s$^{-1}$ Hz$^{-1}$), and
$P(\log L[J_{\rm LW},r],L_{x}[m])\hs d\log$ $J_{\rm LW}=P(\log
L,m)d\log L$, for which the expression was given in
equation~(\ref{eq:uvpdf}).

\begin{figure}
\vbox{\centerline{\epsfig{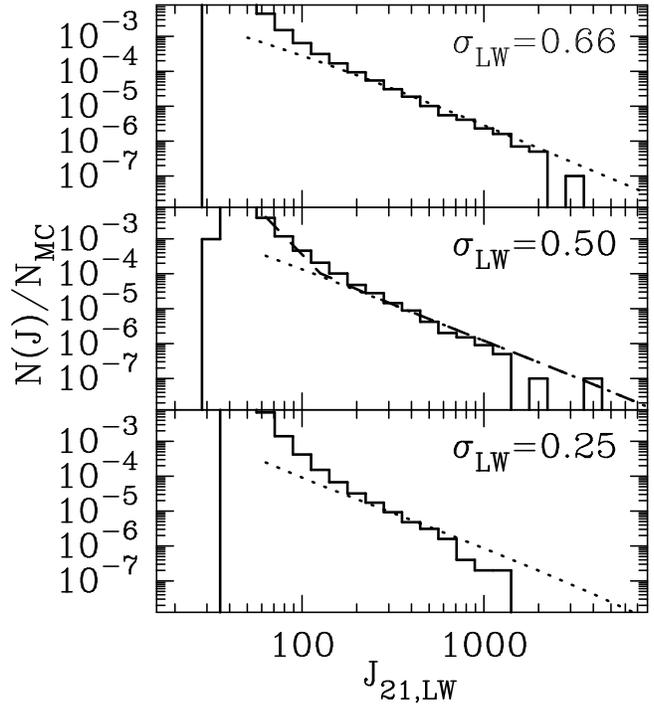}}}
\caption[]{The probability $P_1(J_{\rm LW})$ for a halo to have a
  neighbor that would alone expose it to a Lyman-Werner flux of
  $J_{\rm LW}$ is shown by the {\it dotted curves} (obtained from
  eq.~\ref{eq:analytic}), and compared to the flux PDF obtained from
  Monte-Carlo simulations ({\it histograms}). The fiducial model is
  shown in the {\it middle panel}, while in the {\it top panel} ({\it
    bottom panel}) the luminosity scatter $\sigma_{\rm LW}$ is
  increased (decreased) to $\sigma_{\rm LW}=0.66$ ($\sigma_{\rm
    LW}=0.25$). Both methods yield consistent results for $J_{\rm
    LW,21}\gsim 500$ for all models, demonstrating that the
  high-$J_{\rm 21,LW}$ tail of the $J_{\rm LW}$ distribution is
  dominated by a single nearby star forming halo, and that the
  analytic approach can be used to extrapolate the probability
  distribution of $J_{\rm LW}$ beyond $J_{\rm 21,LW}\gsim 500$. The
  {\it dashed curve} in the middle panel shows a variation of the
  analytical prediction that accounts for {\it pairs} of neighboring
  halos producing the combined flux $J_{\rm LW}$. The agreement at
  lower values of $J_{\rm 21,LW}$ is better, indicating that at
  $J_{\rm LW,21}\lsim 500$, often two halos contribute significantly
  to the total flux.}

\label{fig:checkan}
\end{figure} 
\begin{figure}
\vbox{\centerline{\epsfig{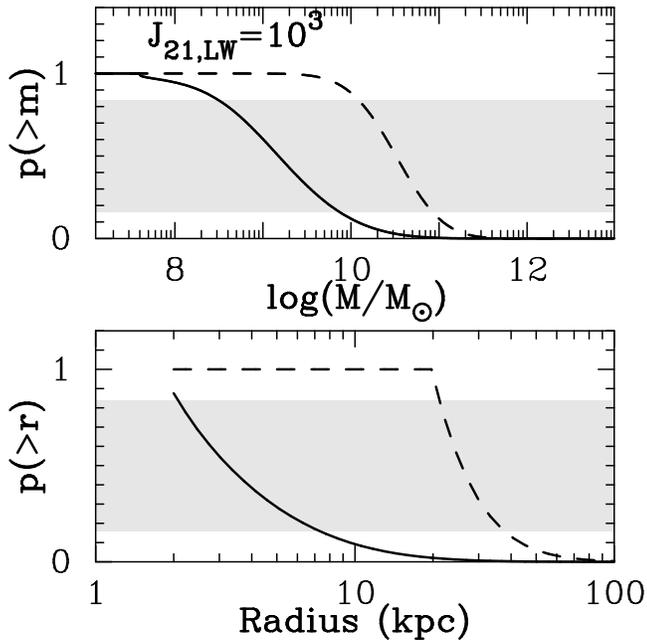}}}
\caption[]{The cumulative probabilities for the mass ({\it top panel})
  and location ({\it bottom panel}) of the single halo that produces
  $J_{\rm 21,LW}=10^{3}$ in the fiducial model ({\it solid curves})
  and for a model in which $r_{\rm min}=20$ kpc ({\it dashed curves
  }). The {\it light grey regions} denote the central 68\%, or $1\sigma$, confidence interval. The figure shows that in the fiducial model, the most probable cases are low--mass halos (in the range $3\times
  10^8M_{\odot}<m<6\times 10^9M_{\odot}$) located very nearby
  ($2<r/{\rm kpc}<7$). The halo mass and distance are both boosted for the model in which $r_{\rm min}$ is increased to 20 kpc.}
\label{fig:pdfs}
\end{figure}

In Figure~\ref{fig:checkan}, we compare the probability $P_1(J_{\rm
  LW})d\log J_{\rm LW}$ obtained from equation~(\ref{eq:analytic})
({\it dotted curves}) with that obtained from the Monte-Carlo
simulations ({\it histograms}) for the fiducial model ({\it middle
  panel}), and for a model in which $\sigma_{\rm LW}$ is changed to
$\sigma_{\rm LW}=0.66$ ($\sigma_{\rm LW}=0.25$) in the {\it top}
({\it bottom}) {\it panel}. Figure~\ref{fig:checkan} shows that the two
methods yield consistent results for $J_{\rm LW,21}\gsim 500$ in all
three cases. This agreement demonstrates that the high fluxes must be
dominated by a single nearby source, and that we can safely use
equation~(\ref{eq:analytic}) to compute the $J_{\rm LW}$-probability
distribution for $J_{\rm 21,LW}\gsim 500$.  At lower values of $J_{\rm
  21,LW}$, the Poisson probability of having more than one nearby
UV-bright star forming halo can become non--negligible.  To examine
this possibility, we define the probability that a halo has {\it
  either} at least one neighbor that would alone expose it to a
Lyman-Werner flux in the range $\log J_{\rm LW}\pm d \log J_{\rm
  LW}/2$ (as before), {\it or } it has at least {\it one pair} of
neighbors whose combined flux equals $\log J_{\rm LW}\pm d \log J_{\rm
  LW}/2$.  In the latter case, the two halos are required to produce
intensities of $yJ_{\rm LW}$ and $[1-y]J_{\rm LW}$, respectively, with
$0<y<1$. 
For a given fractional allocation $y$ of the flux between the
two halos, $P_2(J_{\rm LW})d\log J_{\rm LW}$ is given by $P_2(y)
\propto \{P_1(yJ_{\rm LW})\times P_1([1-y]J_{\rm LW})\}$, and the
total probability for at least one such pair of neighbors is given by
$P_{2,{\rm tot}}=\int_0^1 dy P_2(y)$.
As an example, the {\it dashed curve} in the {\it middle panel} of
Figure~\ref{fig:checkan} shows the probability $P_1+P_{2,{\rm tot}}$
as a function of $J_{\rm LW}$.  Clearly, accounting for pairs of halos
provides a better fit to the MC-simulations at lower values of $J_{\rm 21,LW}$.

In Figure~\ref{fig:pdfs}, we investigate the most likely host halo
mass and distance of the nearby UV source producing the critical
flux. In the {\it top panel}, we plot the cumulative probabilities for
the mass of the halo that produces $J_{\rm 21,LW}=10^{3}$ in the
fiducial model ({\it solid curve}), and for a model in which $r_{\rm
  min}=20$ kpc ({\it dashed curve}). This cumulative probability 
is given by replacing $m_{\rm min}\rightarrow m$ as the lower limit
in the mass integral in equation~(\ref{eq:analytic}).
The {\it solid curve} shows that in the fiducial model, the halo mass lies in the
range $3\times 10^8M_{\odot}<m<6\times 10^9M_{\odot}$, while its most
likely position is $2<r/{\rm kpc}<7$ (the shaded ranges enclose $68\%$ of the
total probability around $p(>x)=0.50$). The most likely halo mass is increased to lie in the range $\approx 10^{10}M_{\odot}<m<8\times 10^{10}M_{\odot}$ for the model in which $r_{\rm min}=20$ kpc. 

In the {\it bottom panel}, we plot the cumulative distribution that
the star forming halo was located at a distance that exceeds $r$
(given by replacing $r_{\rm min}\rightarrow r$ as the lower limit in
the radial integral in eq.~\ref{eq:analytic}).  In the fiducial model,
the distance to the dominant star forming halo lies between $2<r/{\rm
  kpc}<7$, while this range is increased to $20<r/{\rm kpc}<40<$ for
the model in which $r_{\rm min}=20$ kpc ($68\%$--ranges are shaded, as above).

\begin{figure}
\vbox{\centerline{\epsfig{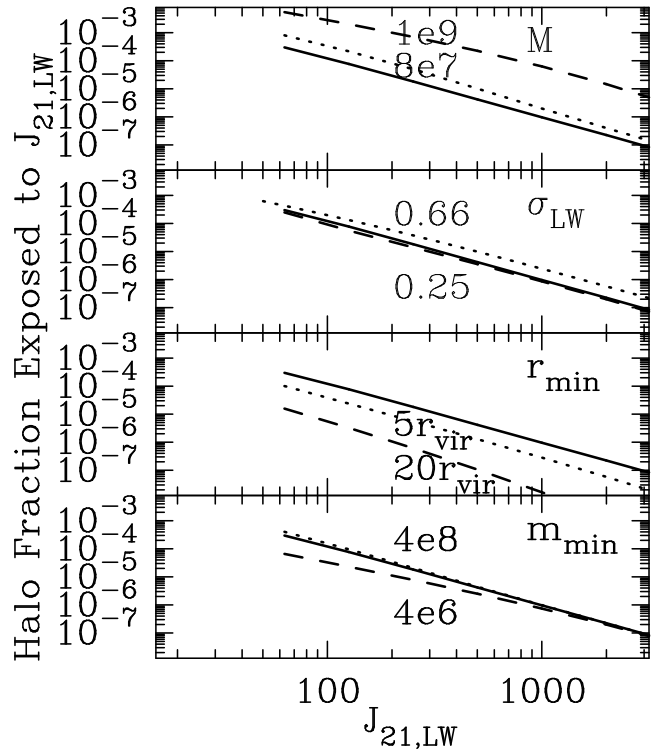}}}
\caption[]{The impact of varying the model parameters, $M$ ({\it top panel}), $\sigma_{\rm LW}$ ({\it second panel from the top}), $r_{\rm min}$ ({\it third panel}), and $m_{\rm min}$ ({\it fourth panel}) on the $J_{\rm LW}$ probability distribution is shown. This figure demonstrates that our results are most sensitive to the precise values of $M$ and $r_{\rm min}$. Increasing $M$ ($r_{\rm min}$) significantly boosts (reduces) the high-flux tail. However, even an increase of $r_{\rm min}$ by a factor of $10$ results in a non-negligible fraction of halos that are exposed to a $J_{\rm 21,LW}=J_{\rm crit}=10^3$ (see text).}
\label{fig:modelvar}
\end{figure}

It is interesting to point out that in the fiducial model, the parameter combination $r=3$ kpc and $m=10^9M_{\odot}$ (for which $p(>r)=p(>m)=0.5$), results in $J_{\rm 21,LW}\sim 200$. Hence, in order for such a halo to produce a local flux of $J_{\rm 21,LW}\sim 10^3$, it must be $\sim 5$ times brighter than average. Since this corresponds to $\sim 1.4\sigma_{\rm LW}$, this implies we are not very sensitive to the tail-end of the assumed log--normal distribution of the UV-luminosity of star forming halos (indeed, in Fig.~\ref{fig:modelvar} we explicitly show that the tail of the flux PDF is not sensitive to $\sigma_{\rm LW}$). 
  
In Figure~\ref{fig:modelvar}, we explore the impact of varying the
model parameters, $M$ ({\it top panel}), $\sigma_{\rm LW}$ ({\it second panel from the top}), as shown above in Fig.~\ref{fig:checkan}), $r_{\rm min}$ ({\it third panel}), and $m_{\rm min}$ ({\it fourth panel}). The parameter dependencies seen in this figure are easily understood.  

Figure~\ref{fig:modelvar} shows that: ({\it i}) the tail end of the flux PDF is boosted slightly for $M=8\times 10^7M_{\odot}$ ({\it dotted curve}) and significantly for $M=10^9M_{\odot}$ ({\it dashed curve}), which reflects the fact that sources are clustered more strongly around more massive halos; ({\it ii}) increasing $\sigma_{\rm LW}$ enhances the tail of the flux PDF, but only slightly. The enhancement arises because increasing $\sigma_{\rm LW}$ enhances the probability of having a nearby halo that is (significantly) brighter than average in the UV\footnote{Consider a halo of mass $m=4\times 10^7 M_{\odot}$ that is located at $r=2$ kpc. In the fiducial model, this source needs to be $\gsim 60$ times brighter than average to produce $J_{\rm LW,21}\gsim 10^3$, which corresponds to $\gsim 3.6 \sigma$. On the other hand, this same source needs to be $\sim 100$ times brighter than average in the model in which $\sigma_{\rm LW}=0.66$ (because the factor $\mathcal{F}$ scales the mean UV luminosity to remain consistent with the observed $z=6$ luminosity function of LBGs, see Table~\ref{table:sigma}), which in this model corresponds to $\gsim 3.1\sigma$. Boosting $\sigma_{\rm LW}$ therefore boosts the number of sources capable of producing $J_{\rm LW,21}\gsim 10^3$ (but only slightly).}; ({\it iii}) increasing $r_{\rm min}$ decreases the tail at high $J_{\rm LW}$. This is because increasing $r_{\rm min}$ eliminates the contribution of very nearby halos to the
$J_{\rm LW}$ distribution (this conclusion can already be inferred
from Fig.~\ref{fig:pdfs}, which shows that the most nearby halos
dominate the single--halo flux); ({\it iv}) increasing $m_{\rm min}$ reduces the flux PDF at $J_{\rm 21,LW}\lsim 500$, but the high--flux tail is not significantly
affected. This is because the highest values of $J_{\rm 21,LW}$
require Lyman-Werner luminosities that are $L=6\times 10^{27}(r_{\rm
  min}/2\hs{\rm kpc})^2(J_{\rm 21,LW}/10^3)$ erg s$^{-1}$
Hz$^{-1}$. Hence, producing these fluxes requires the presence of a
source that is forming stars at a minimum rate of $\sim 0.7M_{\odot}$
yr$^{-1}$. Prolonged star formation rates of this magnitude can only
occur in halos that are more massive than $2\times
10^8M_{\odot}$. Therefore, any variation of $m_{\rm min}$ has a
negligible impact on the flux PDF at $J_{\rm 21,LW}\gsim 500$, as long
as $m_{\rm min}\lsim 2\times 10^8M_{\odot}$ (which seems likely prior to the completion of reionization, e.g. Mesinger \& Dijkstra 2008).

\section{Discussion}

\label{sec:discuss}
\subsection{Lyman-Werner Radiative Transfer}
\label{sec:lwrt}
\begin{figure}
\vbox{\centerline{\epsfig{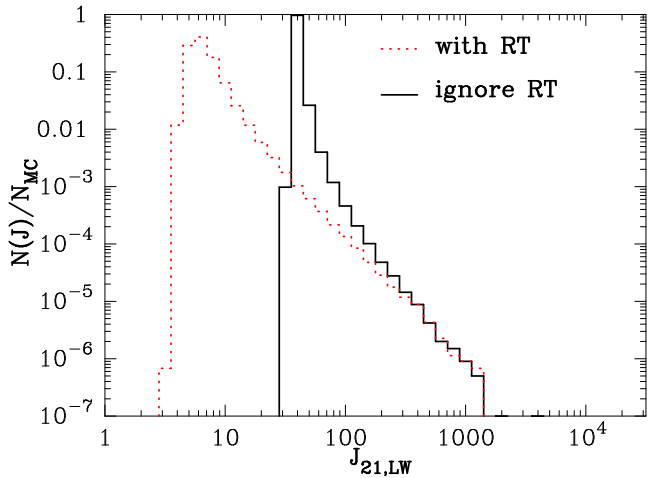}}}
\caption[]{We show the LW flux PDF for the fiducial model ({\it solid black histogram}, as in Fig.~\ref{fig:pdf}), and for a model in which intergalactic H$_2$ reduces the photodissociation rate due to far-away sources, which in turn reduces the value of the global mean LW background by about an order of magnitude. However, the high-$J_{\rm LW}$ tail of the flux PDF depends entirely on the statistics and nature of
very nearby star--forming halos, and LW--radiative transfer effects from the IGM do not affect this tail of the $J_{\rm LW}$ probability distribution at all.}
\label{fig:lywrt}
\end{figure} 
The dissociation of H$_2$ molecules follows photoexcitation of their
electronic states. There are $76$ of these Lyman--Werner transitions,
the majority of them having energies that lie blueward of the
Ly$\beta$ resonance \citep[see, e.g., Fig.~1 of][]{Haiman00}. The
majority of photons that are capable of dissociating H$_2$ molecules
have therefore redshifted into the Ly$\beta$ resonance over a distance
that is much smaller than the $r_{\rm max}$ that was used in our
model. In fact, $r_{\rm max}$ should have a strong frequency
dependence (for example, some of the LW lines are just on the red side
of an atomic H Lyman line; photons could not have suffered significant
redshift before reaching such frequencies without scattering, implying
small $r_{\rm max}$ at the frequency of these LW lines).  

The overall conclusion is that the total photodissociation rate due to far-away sources is reduced, and that our simple modeling which ignores radiate
transfer in the IGM overestimates the value of the global mean
LW-background. Additionally, a trace fraction of molecular hydrogen in
the IGM ($x_{\rm H_2}\sim 2\times 10^{-6}$) is sufficient to produce
an effective optical depth in the LW lines of $\tau_{H_2}\sim 1-2$
\citep{Ricotti01}. Furthermore, $x_{\rm H_2}$ may be enhanced inside ``relic H2 regions'', which increases the volume-averaged effective optical depth of the IGM in the LW lines \citep[][also see Shapiro \& Kang 1987]{J07}. \citet{Kuhlen05} have shown recently that X-rays produced by the first generation of accreting black holes may boost the fraction of intergalactic H$_2$ by a factor of $\sim 20$ around such ``miniquasars''. If the IGM is optically thick in the LW lines, then $r_{\rm max}$ becomes an even stronger function of frequency.

We next investigate the assumption that the IGM is optically thick in
every one of the ${\rm H2}$ lines. Note that this is a very
conservative assumption, requiring that the photo--dissociated
intergalactic ${\rm H2}$ is continuously replenished, which is
unlikely to be the case (Haiman et al. 2000); we purse this
assumption here because it maximizes the impact of RT effects.

In Figure~\ref{fig:lywrt} the {\it red-dotted histogram} shows the LW-flux PDF that is obtained when taking into account this strong frequency dependence of $r_{\rm max}$. We modeled this frequency dependence by computing the maximum distance $r_{\rm max}(k)$ from which LW[k] photons (the k$^{\rm th}$ LW transition, when the 76 LW lines are ranked simply in order of increasing energy) can be received, that is, $Hr_{\rm max}(k)/c=(\lambda_{k}-\lambda_{k+1})/\lambda_{k+1}$. Here, $\lambda_k$ denotes the wavelength of the k$^{\rm th}$ LW transition. LW[k] photons cannot by received from sources at larger radii, as these photons redshifted through the LW[k+1] resonance first, which makes the IGM effectively optically thick to sources beyond $r_{\rm max}(k)$. A source at radius $r$ then contributes the following flux to the local LW-background: $J_{\rm LW}=\frac{1}{4\pi}\frac{L_{\rm UV}}{4\pi r^2}\frac{k_{\rm diss}}{k_{\rm tot}}$, where $k_{\rm tot}$ is the total photodissociation rate as summed over all 76-transitions, and where $k_{\rm diss}$ is the photodissociation rate that one obtains if one sums over only over those lines for which $r<r_{\rm max}(k)$.

Figure~\ref{fig:lywrt} shows that in the above, conservative, optically thick limit, our neglect of
LW radiative transfer overestimates the value of the mean
LW-background by about an order of magnitude, and underestimates the width of the flux PDF. 
However, the high-$J_{\rm LW}$
tail of the flux PDF depends entirely on the statistics and nature of
very nearby star--forming halos (within a few to a few tens of kpc).
LW--radiative transfer effects from the IGM do not affect this tail of
the $J_{\rm LW}$ probability distribution at all, even in this extreme limit.

We also note that LW photons dominate the ${\rm H_2}$--dissociation
rate when the medium is optically thick and lacks radiation above
13.6eV. Indeed, the critical value of $J_{\rm crit}\sim 10^3$
corresponds to the photo--dissociation time {\it due to LW radiation}
being shorter than the free--fall time of the gas.  In our case, the
column density of neutral HI out to the nearby star--forming neighbor
is likely negligible, and direct photo--dissociation by photons with
energies $E>15.4$ eV will boost the total ${\rm H_2}$--dissociation
rate (the latter dominates for neutral HI column densities $N_{\rm
  HI}\lsim 10^{19} {\rm cm^{-2}}$; \citealt{H97}).  This may help to
keep the halo gas ${H_2}$-free, at least during the early stages of
its collapse (until contraction results in a column density exceeding
$N_{\rm HI}\sim 10^{19} {\rm cm^{-2}}$).

Lastly, it is worth pointing out that since the tail of the flux PDF is due to a single neighbor, the impinging LW radiation field is highly anisotropic. The impact of an anisotropic radiation field on the gas could be quite different from that of an isotropic background (e.g., perhaps the 'far side' can be shielded more efficiently against the impinging photodissociating flux, which could allow gas to cool and fragment), but a quantitative assessment of this 3D behavior is beyond the scope of this paper.

\subsection{The Evolutionary Phase of the Halo/Timing Issues}
\label{sec:evo}

We have found that a fraction of a few $\times 10^{-7}$ of all $z=10$
dark matter halos with $T_{\rm vir}\geq10^4$ K are irradiated by a LW
flux that exceeds $J_{\rm crit}$ (Fig.~\ref{fig:modelvar}). However,
the direct--collapse black hole scenario requires that the gas is
irradiated and kept ${\rm H_2}$--free for the entire $\sim$free--fall
time while it collapses to the center of the dark matter halo. This
additional requirement reduces the fraction of dark matter halos in
which these black holes may form.

To quantify this reduction, we identify three relevant time-scales: 
(i) the duration over which halos are producing UV luminosity ($t_{\rm lum}$). In our model, this corresponds to $t_{\rm lum}=\epsilon_{\rm DC}t_{\rm hub}$; (ii) the duration over which the irradiation is needed to avoid fragmentation ($t_{\rm frag}$); and (iii) the spread in formation times of neighboring halos ($t_{\rm form}$). In our model, we implicitly assumed that $t_{\rm form}=t_{\rm hub}$ (since we only assumed that a fraction $\epsilon_{\rm DC}$ of all halos are actively forming stars). The overall likelihood for a combination of two halos in which one prohibits fragmentation in the other is given by the fraction of occasions in which one halo formed within a time $t_{\rm lum}-t_{\rm frag}$ prior to the other during a time $t_{\rm form}$. If the formation times of both halos are uncorrelated, then this probability is given by $q=(t_{\rm lum}-t_{\rm frag})/t_{\rm form}$. If we write $t_{\rm frag}\equiv t_{\rm lum}/f$, then we find that $q=(1-1/f)$. In this case, the reduction due to the synchronization requirement is small for $f\gsim 2$. Furthermore, the reduction is significantly less, if the formation times of the two halos are correlated (which is very likely, since both the collapsing and the star--forming halo formed in the same large--scale overdense region of the Universe).

\subsection{The Importance of Photoheating}
\label{sec:photo}

The halos of interest that are irradiated by a large LW-flux are also illuminated by ionizing radiation, with an ionizing flux that is lower by a factor of $f_{\rm esc}$ (the escape fraction of ionizing photons) than $J_{\rm LW}$.  This escape fraction
is highly uncertain: recently, \citet{Chen07} have derived a distribution function for $f_{\rm esc}$ from the cumulative HI-column density distribution along sightlines to (long-duration) $\gamma$-ray burst (GRB) afterglows. Long-duration GRBs are thought to be associated with massive ($M \gsim 10-30 M_{\odot}$) young stars, and are therefore thought to occur within star forming regions of galaxies (see e.g. Woosley \& Bloom 2006). GRB afterglows may therefore probe the HI column-density along sightlines to the production sites of ionizing photons\footnote{Furthermore, evidence exists that the majority of GRBs occur in sub-$L_{*}$ galaxies (see Chen et al. 2007, and references therein), which are likely to bear a closer resemblance to the $z=10$ starforming galaxies in our model, than the massive LBGs for which escape fractions of ionizing photons have also been determined on the basis of their observed Lyman Break \citep{S06}.}. \citet{Chen07} show that $f_{\rm esc}$ may obtain a wide range of values, and that $\langle f_{\rm esc}\rangle=0.02$, in good agreement with theoretical calculations of the escape fraction of ionizing photons from galaxies in adaptive mesh refinement hydrodynamical simulations\footnote{\citet{Whalen04} found $f_{\rm esc}\sim 1$ for galaxies at $z\sim 20$, using one-dimensional hydrodynamic calculation. However, their calculations focused on the first UV-sources in the Universe which populated much less massive ``minihalos''.} by \citet{Gnedin08}.

From the above we conclude that the halos of interest may be irradiated by a ionizing flux that likely has a broad distribution between $J_{\rm ion,21}=0-10^3$, with $\langle J_{\rm ion,21}\rangle\sim 20$. To investigate the impact of this photoionizing flux on the dynamics of the gas in collapsing halo, we performed a 1-D hydrodynamical simulation with a code that was originally written by \citet{TW95} and modified as described in Dijkstra et al. (2004, where it was used to study the impact of photoionization feedback on the formation of dwarf galaxies). We illuminated a gas cloud of mass $M=10^8M_{\odot}$ with an ionizing flux having amplitudes of $J_{\rm ion,21}=10$ and $J_{\rm ion,21}=10^2$, and spectral index $\alpha=1$ (i.e. the ionizing flux density is given by $J(\nu)=J_{\rm 21,ion}(\nu/\nu_L)^{-\alpha}\times 10^{-21}$ erg s$^{-1}$ Hz$^{-1}$ cm$^{-2}$ sr$^{-1}$). This ionizing flux was switched on at a time $\Delta t=\epsilon_{\rm DC}t_{\rm hub}$ before the cloud would have collapsed to $r=0$ in the absence of pressure. 
\begin{figure}
\vbox{\centerline{\epsfig{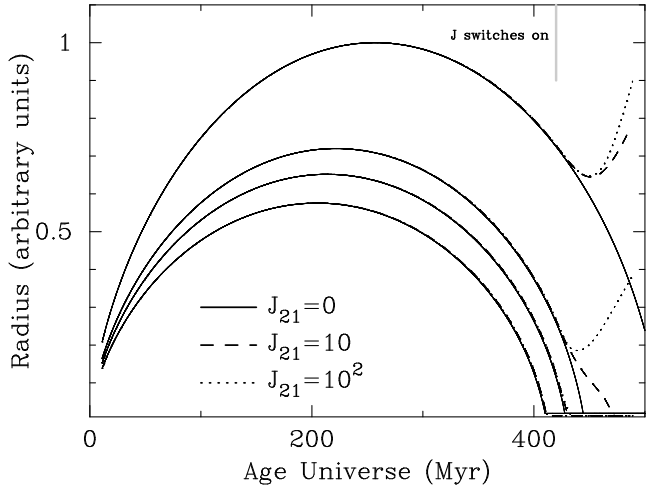}}}
\caption[]{{\it Solid curves} show the time evolution of the radii of gas shells during the collapse of $10^8M_{\odot}$ halo, in the absence of an externally generated ionizing radiation field (i.e $J_{\rm ion,21}=0$). Each set of curves corresponds to shells (initially) enclosing 30\%, 40\%, 50\% and 100\% of the total halo mass. All gas in the halo has collapsed by $z=10$, when $t_{\rm hub}\sim 500$ Myr. The {\it dashed} ({\it dotted curves}) represent models in which an ionizing background, whose amplitude was $J_{\rm ion,21}=10$ ($J_{\rm ion,21}=10^2$), was switched at a time $\Delta t=\epsilon_{\rm DC}t_{\rm hub}$ before the cloud would have collapsed to $r=0$ in the absence of pressure. The figure shows that even when $J_{\rm ion,21}=10$ ($J_{\rm ion,21}=10^2$), still $50\%$ ($40\%$) of the gas was able to collapse.}
\label{fig:photoion}
\end{figure} 

The result of these calculations are shown in Figure~\ref{fig:photoion}. {\it Solid curves} show the time evolution of the radii of gas shells enclosing 30\%, 40\%, 50\% and 100\% of the total halo mass in the absence of an externally generated ionizing radiation field. The {\it dashed} ({\it dotted curves}) represent the models in which $J_{\rm ion,21}=10$ ($J_{\rm ion,21}=10^2$). The figure shows that even when $J_{\rm ion,21}=10$ ($J_{\rm ion,21}=10^2$), still $50\%$ ($40\%$) of the gas was able to collapse, which implies that photoionization feedback does not prevent the gas from collapsing, even with extremely high UV fluxes. The main reason for this is that photoionization feedback is most efficient when it can operate for a prolonged time: e.g. in our spherical collapse model, this feedback mechanism has the largest effect when the radiation field is switched 'on'  when (or before) the gas completely decouples from the Hubble flow and starts to collapse. However, in the final $\epsilon_{\rm DC}t_{\rm hub}$ of the cloud's collapse, the gas in the halo has already collapsed to densities that greatly exceed the mean cosmic gas density in the Universe, and cooling processes are more efficient. Furthermore, at these high gas densities self-shielding, which is ignored in our simulation, will become important which would make photoionization feedback even less important. 

Lastly, it is worth pointing out that halos form, and cluster strongly, at intersections of filaments. The dominating LW-flux will therefore be emitted by halos along the filaments. On the other hand, their ionizing flux will most likely escape into the low density regions, perpendicular to the filaments. This argues for a value of $J_{\rm ion,21}/J_{\rm 21,LW}$ that is lower than $\langle f_{\rm esc}\rangle$. Furthermore, the simulations by \citet{Gnedin08} suggest that the escape fraction of ionizing photons varies significantly from sightline-to-sightline, with a small fraction of sightlines having $f_{\rm esc}=1$, and a significant fraction having $f_{\rm esc} \lsim 10^{-3}$ (mainly because of intervening HI and HeI). This also implies that it is likely that $J_{\rm ion,21}/J_{\rm 21,LW}\lsim \langle f_{\rm esc} \rangle$.

\subsection{The Transition from Massive to Supermassive Black Hole}
\label{sec:final}

Even if a $10^4-10^6M_{\odot}$ black hole forms as a direct result of
the collapse of a gas cloud that was irradiated by a LW flux that
exceeded $J_{\rm crit}$, the black hole still needs to grow to
$10^9M_{\odot}$. This requires the black hole to consume a tremendous
amount of gas over $\sim 0.5$ Gyr (the total time that elapsed between
$z=10$ and $z=6$). This implies that the original host halo in
which the black hole formed, must merge with other halos into a
massive $10^{12}-10^{13}M_{\odot}$ host halo (which also agrees with the
mass of the host halos of the $z\sim 6$ quasars; see, e.g., the review
by Haiman \& Quataert 2004). Interestingly, we found that those $M=10^8M_{\odot}$ halos that are exposed to a LW
flux exceeding $J_{\rm crit}$, are the ones with a nearby star
forming galaxy. The halo mass of this nearby galaxy is possibly as
large as $10^{11}M_{\odot}$ (see \S~\ref{sec:param}), and the
$10^4-10^6M_{\odot}$ black hole is likely to end up inside this larger
dark matter halo.

Utilizing \citet[][eq. 2.16]{Lacey93}, we find that the fraction of
all $10^{8}M_{\odot}$ halos at $z=10$ that are incorporated into
$M\geq 10^{12}M_{\odot}$ halos by $z=6$ is $1.4\times 10^{-4}$.  This
fraction is much larger (by a factor of $\sim 10^3$) than the fraction
of halos that have close and massive neighbors and are irradiated by a
critical flux. This is strongly suggestive that essentially all of
these strongly irradiated halos are in the highly clustered tail of
the spatial distribution, and will end up in massive halos, such as
those hosting the SDSS quasars, by $z=6$. For reference, we also note
that the $10^{8}M_{\odot}$ halos contain a fraction $1.6\times
10^{-2}$ of the total mass at $z=10$, while $10^{12}M_{\odot}$ halos
at $z=6$ have a mass--fraction of $2.8\times 10^{-5}$ at $z=6$.  This
implies that $\approx (1.6\times 10^{-2})\times (1.4\times 10^{-4})/(2.8\times10^{-5})\approx 10\%$ of the masses of the $10^{12}M_{\odot}$
halos at $z=6$ were already assembled into $10^{8}M_{\odot}$ halos at
$z=10$. In this picture, the $z=6$ supermassive black holes
originally formed inside collapsing $10^8M_{\odot}$ satellite halos of
more massive star forming galaxies.

\subsection{Metal Enrichment of $T_{\rm vir}> 10^4$ K Halos}

While the above discussion shows that a non--negligible fraction of halos may be kept ${\rm H_2}$--free by a strong LW background, another possible
effect that could invalidate the direct--collapse black hole formation
scenario is the presence of heavy elements in such halos. Indeed,
most high--redshift DM halos with $T_{\rm vir}\sim 10^4$K may be
already enriched with at least trace amounts of metals and dust
produced by prior star--formation in their progenitors.
\citet{Omukai08} recently studied the thermal and chemical evolution
of low--metallicity gas exposed to extremely strong UV radiation
fields. They find that these metals and any accompanying dust can
catastrophically lower the gas temperature, producing an effective
equation--of--state index of $\gamma\equiv d\ln p/d\ln\rho \ll 1$,
even in the presence of a UV background as strong as $J_{\rm LW}=10^3$,
which keeps ${\rm H_2}$ photodissociated. They suggest that gas
fragmentation may thus be inevitable above a critical metallicity,
whose value is between $Z_{\rm cr} \approx 3 \times 10^{-4} Z_{\odot}$
(in the absence of dust) and as low as $Z_{\rm cr} \approx 5 \times
10^{-6} Z_{\odot}$ (with a dust-to-gas mass ratio of about $0.01
Z/Z_{\odot}$, also see Omukai 2000, Bromm et al. 2001, Schneider et al. 2003). When the metallicity exceeds these critical values,
dense clusters of low--mass stars may form at the halo nucleus.  While
relatively massive stars in such a cluster could then rapidly coalesce
into a single more massive object (which may produce an
intermediate--mass BH remnant with a mass up to $M\lsim 10^2-10^3~{\rm
  M_\odot}$), the formation of much BHs as massive as 
$M\lsim 10^5-10^6~{\rm  M_\odot}$  by direct collapse
is likely to require -- in addition to being exposed to a large $J_{\rm crit}$ -- that the halo gas is also essentially metal--free. Metal enrichment by the nearby star forming galaxy (that is responsible for the large local LW flux),  is not necessarily a problem, since the expelled metals most efficiently mix in the outer edge of the collapsing halo, while leaving the inner core pristine \citep[][]{G07,CR08}.

\section{Conclusions}
\label{sec:conc}

H$_2$ molecules can be photo--dissociated by ultra--violet (UV) radiation, either directly (by photons with energies $E>14.7$ eV, if the molecules are exposed to ionizing radiation) or as a result of electronic excitation by
Lyman--Werner (hereafter LW) photons with energies 11.2eV$\lsim$ E
$<E_{\rm H}$. In this paper we focused on this latter process, which operates
 even in gas that is self-shielded, and/or in gas that is shielded by a neutral intergalactic medium (IGM) prior to the completion of reionization (see \citealt{H97}), from radiation at $E>E_{\rm H}$. 

Photodissociation feedback possibly plays an important role in the
formation of the supermassive black holes (SMBHs, $M_{\rm BH}\sim
10^9M_{\odot}$) that existed at redshift $z>6$, when the age of the
universe was $<1$ Gyr \citep[e.g.][]{FanReview}. The reason for this is that this feedback mechanism can affect gas cooling and
collapse into halos that have $T_{\rm vir}> 10^4$ K
(e.g. \citealt{OH02}). Without H$_2$ molecules, gas inside these halos
collapses nearly isothermally due to atomic cooling, with the
temperature remaining as high as $T\sim 10^4$K, which may strongly
suppress the ability of gas to fragment into stellar mass objects
\citep[e.g.][]{OH02}. Instead of fragmenting, this gas could rapidly
accrete onto a seed BH \citep{Volonteri05}, or collapse directly into
a very massive ($M_{\rm BH}=10^4-10^6M_{\odot}$) black hole
\citep{BL03,K04,Begelman06,SS06,Lo06,Lo07}, possibly with an intermediate state in
the form of a very massive star \citep[see][for a more complete
review]{Omukai08}. If these ``direct-collapse black holes'' indeed
formed, then they would provide a head--start that could help explain
the presence of SMBHs with inferred masses of several $\times 10^9{\rm
  M_\odot}$ by $z\approx 6$.

Possibly the most stringent requirement for the direct-collapse black hole model, is the intensity of the photodissociating background, which must exceed $J_{\rm 21,LW}\geq J_{\rm crit}\sim 10^3$ (Bromm \& Loeb 2003, where $J_{\rm 21,LW}$ denotes the intensity in the LW background in units of $10^{-21}$ erg s$^{-1}$ Hz$^{-1}$ sr$^{-1}$ cm$^{-2}$, evaluated at the Lyman limit). Since the mean free path of Lyman-Werner photons is large, the LW-background is expected to be nearly uniform, and likely significantly lower than $J_{\rm crit}$ \citep{Omukai08}. However, some variation in the LW-flux that irradiates individual dark matter halos is expected. For example, a dark matter halo with a nearby star forming galaxy may 'see' a LW-flux that exceeds the background by orders of magnitude. 

In this paper, we have modeled this variation and we have computed the probability distribution function (PDF) of the LW-flux, $J_{\rm LW}$, that irradiates a high-redshift dark matter halo at $z=10$. The main differences from previous works that study the spatial variation of the UV background are that: (i) we include (1) the non--linear clustering of sources, using the non-linear bias formalism developed by Iliev et al. (2003), which describes the clustering of closely separated halos particularly well; (2) Poisson fluctuations in the number of surrounding star forming galaxies; (3) the probable existence of a dispersion in LW luminosity for a given halo mass.  In the local Universe the UV luminosity can vary by orders of magnitude for a given stellar mass \citep[e.g][]{DS07}, and a similar variation of the UV luminosity may be expected for a given halo mass; (ii) we specialize to compute the PDF of the flux as seen by high--redshift DM halos; (iii) we focus on the tail of the flux PDF, and (iv) we discuss the significance of this tail for rapid high--redshift SMBH growth.

We found that $>99\%$ of the dark matter halos at $z=10$ are illuminated by LW flux that is within a factor of $2$ from the global background value (Fig.~\ref{fig:pdf}). However, a tiny fraction ($f_{\rm crit}=10^{-8}-10^{-6}$) of dark matter halos with virial temperatures exceeding $T_{\rm vir}=10^4$ K may see a LW-flux that is boosted to the level $J_{\rm crit}=10^3$ that is required by the direct collapse black hole model, due to nearby luminous neighbors. We demonstrated that this result is insensitive both to our assumed model parameters (Fig.~\ref{fig:modelvar}), and to radiative transfer effects due to intervening atomic and molecular hydrogen gas (\S~\ref{sec:lwrt}). 

Depending on (i) the duration for which halos are producing UV luminosity ($t_{\rm lum}$), (ii) the duration for which the irradiation is needed to avoid fragmentation ($t_{\rm frag}$), and (iii) the spread in formation times of halos ($t_{\rm form}$), these rare halo pairs provide possible sites in which primordial gas clouds collapse directly into massive black holes ($M_{\rm BH}\approx 10^{4-6}M_{\odot}$, see \S~\ref{sec:evo}). In \S~\ref{sec:final}, we showed that these black holes are likely to end up in massive ($10^{12}-10^{13}M_{\odot}$) host halos (which also agrees with the mass of the host halos of the $z\sim 6$ quasars; see, e.g., the review by Haiman \& Quataert 2004), in which they can grow to $10^9M_{\odot}$ within $\sim 0.5$ Gyr (the total time that elapsed between $z=10$ and $z=6$). It is interesting to stress that in this picture, the $z=6$ supermassive black holes originally formed inside collapsing $10^8M_{\odot}$ satellite halos of more massive star forming galaxies.

The fraction $f_{\rm crit}\sim 10^{-8}-10^{-6}$ translates to $\sim 10-10^3$ $\gsim 10^8M_{\odot}$ halos per comoving Gpc$^3$ volume. The gas inside one of these halos needs to collapse directly into a massive black hole, and subsequently grow to a $M_{\rm BH}=10^9M_{\odot}$ to explain observed space density of $z=6$ luminous quasars. The actual number of supermassive black holes at $z=6$ is higher by a factor of $1/\epsilon_{\rm QSO}$, where $\epsilon_{\rm QSO}$ denotes the fraction of Hubble time that $z=6$ supermassive black holes are observable as luminous quasars. Assuming a quasar lifetime of $\sim 50$ Myr \citep[e.g.][]{Martini04}, we have $\epsilon_{\rm QSO}\sim 0.05$, and the number density of $z=6$ supermassive black holes is $\sim 20$ cGpc$^{-3}$, which we conclude is manageable with the direct-collapse black hole scenarios.

{\bf Acknowledgments} MD is supported by Harvard University funds. ZH
acknowledges support by the Pol\'anyi Program of the Hungarian
National Office of Technology.  JSBW acknowledges the support of the
ARC. Partial support for this work was also provided by NASA through Hubble Fellowship grant \#HF-01222.01 to AM, awarded by the Space Telescope Science Institute, which is operated by the Association of Universities for Research in Astronomy, Inc., for NASA, under contract NAS 5-26555.

\label{lastpage}

\begin{thebibliography}{14}
\expandafter\ifx\csname natexlab\endcsname\relax\def\natexlab#1{#1}\fi
\bibitem[Ahn et al.(2008)]{Ahn08} Ahn, K., Shapiro, P.~R., 
Iliev, I.~T., Mellema, G., 
\& Pen, U.-L.\ 2008, ArXiv e-prints, 807, arXiv:0807.2254 


\bibitem[Barkana \& Loeb(2001)]{Barkana01} Barkana, R., \& Loeb, A.\ 2001, \physrep, 349, 125 

\bibitem[Barkana \& Loeb(2005a)]{BL05} Barkana, R., \& Loeb, 
A.\ 2005a, \apjl, 624, L65 

\bibitem[Barkana \& Loeb(2005b)]{fluc} Barkana, R., \& Loeb, A.\ 2005b, \apj, 626, 1 

\bibitem[Begelman et al.(2006)]{Begelman06} Begelman, M.~C., 
Volonteri, M., \& Rees, M.~J.\ 2006, \mnras, 370, 289 

\bibitem[Bouwens et al.(2006)]{Bouwens06} Bouwens, R.~J., 
Illingworth, G.~D., Blakeslee, J.~P., \& Franx, M.\ 2006, \apj, 653, 53 

\bibitem[Bromley, Somerville \& Fabian(2004)]{Bromley04} Bromley, J. M., Somerville, R.S., \& Fabian, A. C. 2004, MNRAS 350, 456

\bibitem[Bromm et al.(2001)]{Bromm01} Bromm, V., Ferrara, A., 
Coppi, P.~S., \& Larson, R.~B.\ 2001, \mnras, 328, 969 

\bibitem[Bromm \& Loeb(2003)]{BL03} Bromm, V., \& Loeb, A.\ 
2003, \apj, 596, 34 

\bibitem[Bullock et al.(2000)]{Bullock00} Bullock, J.~S., 
Kravtsov, A.~V., \& Weinberg, D.~H.\ 2000, \apj, 539, 517 

\bibitem[Cen(2003)]{Cen03} Cen, R.\ 2003, \apj, 591, 12 

\bibitem[Cen \& Riquelme(2008)]{CR08} Cen, R., \& Riquelme, M.~A.\ 2008, \apj, 674, 644 

\bibitem[Chen et al.(2007)]{Chen07} Chen, H.-W., Prochaska, 
J.~X., \& Gnedin, N.~Y.\ 2007, \apjl, 667, L125 

\bibitem[Ciardi et al.(2000)]{Ciardi00} Ciardi, B., Ferrara, A., 
Governato, F., \& Jenkins, A.\ 2000, \mnras, 314, 611 

\bibitem[Couchman \& Rees(1986)]{CR86} Couchman, H.~M.~P., \& Rees, M.~J.\ 1986, \mnras, 221, 53 

\bibitem[Croft(2004)]{Croft04} Croft, R.~A.~C.\ 2004, \apj, 
610, 642 

\bibitem[Dijkstra et al.(2004)]{feedback} Dijkstra, M., Haiman, 
Z., Rees, M.~J., \& Weinberg, D.~H.\ 2004, \apj, 601, 666 

\bibitem[Dijkstra et al.(2007)]{LF} Dijkstra, M., Wyithe, 
J.~S.~B., \& Haiman, Z.\ 2007, \mnras, 379, 253 

\bibitem[Dunkley et al.(2008)]{wmap5} Dunkley, J., et al. 2008, \apj, submitted, arXiv.org:0803.0586

\bibitem[Efstathiou(1992)]{ef92} Efstathiou, G.\ 1992, 
\mnras, 256, 43P 

\bibitem[Fan et al.(2001)]{fan01} Fan, X., et al.\ 2001, \aj, 
122, 2833 

\bibitem[Fan et al.(2003)]{fan03} Fan, X., et al.\ 2003, \aj, 
125, 1649 
\bibitem[Fan(2006)]{FanReview} Fan, X.\ 2006, New Astronomy 
Review, 50, 665 

\bibitem[Gnedin et al.(2008)]{Gnedin08} Gnedin, N.~Y., Kravtsov, 
A.~V., \& Chen, H.-W.\ 2008, \apj, 672, 765 

\bibitem[Greif et al.(2007)]{G07} Greif, T.~H., Johnson, 
J.~L., Bromm, V., \& Klessen, R.~S.\ 2007, \apj, 670, 1 

\bibitem[Haiman(2004)]{h04} Haiman, Z 2004, ApJ, 613, 36

\bibitem[Haiman \& Loeb(2001)]{hl01} Haiman, Z., \& Loeb, A. 2001, ApJ, 552, 459

\bibitem[Haiman et al.(1996)]{H96} Haiman, Z., Thoul, 
A.~A., \& Loeb, A.\ 1996, \apj, 464, 523 

\bibitem[Haiman et al.(1997)]{H97} Haiman, Z., Rees, M.~J., \& Loeb, A.\ 1997, \apj, 476, 458 

\bibitem[Haiman et al.(2000)]{Haiman00} Haiman, Z., Abel, T., 
\& Rees, M.~J.\ 2000, \apj, 534, 11 

\bibitem[Haiman \& Holder(2003)]{HH03} Haiman, Z., \& 
Holder, G.~P.\ 2003, \apj, 595, 1 

\bibitem[Hennawi et al.(2006)]{Hennawi} Hennawi, J.~F., et al.\ 
2006, \aj, 131, 1 

\bibitem[Iliev et al.(2003)]{Iliev03} Iliev, I.~T., 
Scannapieco, E., Martel, H., \& Shapiro, P.~R.\ 2003, \mnras, 341, 81 

\bibitem[Inoue et al.(2006)]{Inoue06} Inoue, A.~K., Iwata, I., 
\& Deharveng, J.-M.\ 2006, \mnras, 371, L1 

\bibitem[Johnson \& Bromm(2007)]{JB07} Johnson, J.~L., \& Bromm, V.\ 2007, \mnras, 374, 1557 

\bibitem[Johnson et al.(2007)]{J07} Johnson, J.~L., Greif, 
T.~H., \& Bromm, V.\ 2007, \apj, 665, 85 

\bibitem[Johnson et al.(2008)]{J08} Johnson, J.~L., Greif, 
T.~H., \& Bromm, V.\ 2008, \mnras, 388, 26 

\bibitem[Kennicutt(1998)]{K98} Kennicutt, R.~C., Jr.\ 1998, 
\araa, 36, 189 

\bibitem[Kitayama \& Ikeuchi(2000)]{K00} Kitayama, T., \& 
Ikeuchi, S.\ 2000, \apj, 529, 615 

\bibitem[Komatsu et al.(2008)]{wmap} Komatsu, E., et al.\ 2008, \apj, submitted, arXiv.org:0803.0547

\bibitem[Koushiappas et al.(2004)]{K04} Koushiappas, S.~M., 
Bullock, J.~S., \& Dekel, A.\ 2004, \mnras, 354, 292 

\bibitem[Kuhlen \& Madau(2005)]{Kuhlen05} Kuhlen, M., \& Madau, P.\ 2005, \mnras, 363, 1069 

\bibitem[Lacey \& Cole(1993)]{Lacey93} Lacey, C., \& Cole, S.\ 1993, \mnras, 262, 627 

\bibitem[Lee et al.(2008)]{Lee08} Lee, K.~-., Giavalisco, M., 
Conroy, C., Wechsler, R.~H., Ferguson, H.~C., Somerville, R.~S., Dickinson, 
M.~E., \& Urry, C.~M.\ 2008, ArXiv e-prints, 808, arXiv:0808.1727 

\bibitem[Lepp \& Shull(1984)]{LS84} Lepp, S., \& Shull, J.~M.\ 1984, \apj, 280, 465 

\bibitem[Li et al.(2007)]{Li07} Li, Y., et al.\ 2007, \apj, 
665, 187 
\bibitem[Lodato \& Natarajan(2006)]{Lo06} Lodato, G., \& Natarajan, P.\ 2006, \mnras, 371, 1813 

\bibitem[Lodato \& Natarajan(2007)]{Lo07} Lodato, G., \& Natarajan, P.\ 2007, \mnras, 377, L64 

\bibitem[Mesinger et al.(2006)]{M06} Mesinger, A., Bryan, 
G.~L., \& Haiman, Z.\ 2006, \apj, 648, 835 

\bibitem[Meiksin \& White(2004)]{Meiksin04} Meiksin, A., \& White, M.\ 2004, \mnras, 350, 1107 

\bibitem[Mesinger \& Furlanetto(2007)]{MF07} Mesinger, A., \& Furlanetto, S.\ 2007, \apj, 669, 663 

\bibitem[Mesinger \& Furlanetto(2008)]{2008MNRAS.386.1990M} Mesinger, A., \& Furlanetto, S.~R.\ 2008, \mnras, 386, 1990 

\bibitem[Mesinger \& Dijkstra (2008)]{MD08} Mesinger, A., Dijkstra, M., 2008, MNRAS in press.


\bibitem[Martini(2004)]{Martini04} Martini, P.\ 2004 in ``Coevolution of Black Holes and Galaxies'', Carnegie Observatories Astrophysics Series, Vol. 1, Ed. L. C. Ho. (Cambridge, U.K.: Cambridge University Press), p. 169

\bibitem[Miralda-Escud{\'e} et al.(2000)]{MHR} 
Miralda-Escud{\'e}, J., Haehnelt, M., \& Rees, M.~J.\ 2000, \apj, 530, 1 
\bibitem[Mo \& White(1996)]{MoWhite96} Mo, H.~J., \& White, S.~D.~M.\ 1996, \mnras, 282, 347 

\bibitem[Oh \& Haiman(2002)]{OH02} Oh, S.~P., \& Haiman, Z.\ 2002, \apj, 569, 558 

\bibitem[Omukai(2000)]{Omukai00} Omukai, K.\ 2000, \apj, 534, 
809 

\bibitem[Omukai et al.(2008)]{Omukai08} Omukai, K., Schneider, 
R., \& Haiman, Z.\ 2008, ArXiv e-prints, 804, arXiv:0804.3141 


\bibitem[Press \& Schechter(1974)]{PS} Press, W.~H., \& Schechter, P.\ 1974, \apj, 187, 425 

\bibitem[Ricotti et al.(2001)]{Ricotti01} Ricotti, M., Gnedin, 
N.~Y., \& Shull, J.~M.\ 2001, \apj, 560, 580 

\bibitem[Saslaw \& Zipoy(1967)]{SZ67} Saslaw, W.~C., \& Zipoy, D.\ 1967, \nat, 216, 976 

\bibitem[Scannapieco 
\& Barkana(2002)]{SB02} Scannapieco, E., \& Barkana, R.\ 2002, \apj, 571, 585 

\bibitem[Schaerer(2003)]{Sch03} Schaerer, D.\ 2003, \aap, 
397, 527 

\bibitem[Schaye(2006)]{Schaye06} Schaye, J.\ 2006, \apj, 643, 59 

\bibitem[Schiminovich et al.(2007)]{DS07} Schiminovich, D., 
et al.\ 2007, ApJ in press, arXiv:0711.4823 

\bibitem[Schneider et al.(2003)]{S03} Schneider, R., 
Ferrara, A., Salvaterra, R., Omukai, K., \& Bromm, V.\ 2003, \nat, 422, 869 

\bibitem[Shapiro \& Kang(1987)]{SK87} Shapiro, P.~R., \& Kang, H.\ 1987, \apj, 318, 32 

\bibitem[Shapiro(2005)]{shapiro05} Shapiro, S. L. 2005, ApJ, 620, 59

\bibitem[Shapley et al.(2006)]{S06} Shapley, A.~E., 
Steidel, C.~C., Pettini, M., Adelberger, K.~L., \& Erb, D.~K.\ 2006, \apj, 
651, 688 

\bibitem[Sheth et al.(2001)]{ST01} Sheth, R.~K., Mo, H.~J., 
\& Tormen, G.\ 2001, \mnras, 323, 1 

\bibitem[Siana et al.(2007)]{Siana07} Siana, B., et al.\ 2007, 
\apj, 668, 62 

\bibitem[Spaans \& Silk(2006)]{SS06} Spaans, M., \& Silk, J.\ 2006, \apj, 652, 902 

\bibitem[Stark et al.(2007)]{Stark07} Stark, D.~P., Ellis, 
R.~S., Richard, J., Kneib, J.-P., Smith, G.~P., 
\& Santos, M.~R.\ 2007, \apj, 663, 10 

\bibitem[Tanaka \& Haiman(2008)]{Taka08} Tanaka, T., \& Haiman, Z. 2008, ApJ, submitted

\bibitem[Thoul \& Weinberg(1995)]{TW95} Thoul, A.~A.~\& 
Weinberg, D.~H.\ 1995, \apj, 442, 480 

\bibitem[Thoul \& Weinberg(1996)]{tw96} Thoul, A.~A., \& 
Weinberg, D.~H.\ 1996, \apj, 465, 608 

\bibitem[Volonteri \& Rees(2005)]{Volonteri05} Volonteri, M., \& Rees, M.~J.\ 2005, \apj, 633, 624 

\bibitem[Volonteri \& Rees(2006)]{Volonteri06} Volonteri, M., \& Rees, M.~J.\ 2006, \apj, 650, 669 

\bibitem[Volonteri et al.(2008)]{V08} Volonteri, M., 
Lodato, G., \& Natarajan, P.\ 2008, \mnras, 383, 1079 

\bibitem[Whalen et al.(2004)]{Whalen04} Whalen, D., Abel, T., 
\& Norman, M.~L.\ 2004, \apj, 610, 14 

\bibitem[Woosley \& Bloom(2006)]{GRB} Woosley, S.~E., \& Bloom, J.~S.\ 2006, \araa, 44, 507 

\bibitem[Wyithe \& Loeb(2003)]{WL03} Wyithe, J.~S.~B., \& 
Loeb, A.\ 2003, \apj, 586, 693 

\bibitem[Wyithe \& Loeb(2005)]{WL05} Wyithe, J.~S.~B., \& 
Loeb, A.\ 2005, \apj, 625, 1 

\bibitem[Wyithe \& Loeb(2006)]{WL06} Wyithe, J.~S.~B., \& 
Loeb, A.\ 2006, \nat, 441, 322 

\bibitem[Yoo \& Miralda-Escud\'e(2004)]{ym04} Yoo, J., \& Miralda-Escud\'e, J. 2004, ApJ, 614, 25

\bibitem[Zuo(1992)]{Zuo92} Zuo, L.\ 1992, \mnras, 258, 36 
\end{thebibliography}
\end{document}